\documentclass[11pt,fleqn]{article}
  \usepackage{amsmath,amssymb,graphicx,epsfig}
 \usepackage{caption}

  \numberwithin{equation}{section}
  \usepackage{color}

  \usepackage{cancel}

  \newcommand{\Tr}{\text{Tr}}
  
  \newcommand{\re}{\text{Re}}

\begin{document}
\begin{flushright}
CERN-PH-TH-2015-170\\
\end{flushright}

\thispagestyle{empty}

\vspace{2cm}

\begin{center}
{\Large {\bf 
MSSM soft terms from supergravity with gauged R-symmetry in de Sitter vacuum
}}
\\
\medskip
\vspace{1.cm}
\textbf
{
I. Antoniadis$^{\,a,b}$, 
R. Knoops$^{\, c,d,e}$}
\bigskip

$^a$ {\small LPTHE, UMR CNRS 7589
Sorbonne Universit\'es, UPMC Paris 6, 75005 Paris France}

$^b$ {\small Albert Einstein Center, Institute for Theoretical Physics
Bern University, Sidlerstrasse 5, CH-3012 Bern, Switzerland }

$^c$ {\small CERN Theory Division, CH-1211 Geneva 23, Switzerland}

$^d$ {\small Section de Math\'ematiques, Universit\'e de Gen\`eve, CH-1211 Gen\`eve, Switzerland}

$^e$ {\small Instituut voor Theoretische Fysica, KU Leuven,
Celestijnenlaan 200D, B-3001 Leuven, Belgium}

\end{center}
\bigskip

  \setcounter{page}{1}
  \begin{abstract}
  
  We work out the phenomenology of a model of supersymmetry breaking in the presence of a tiny (tunable) positive cosmological constant, proposed by the authors in arXiv:1403.1534. It utilises a single chiral multiplet with a gauged shift symmetry, that can be identified with the string dilaton (or an appropriate compactification modulus). The model is coupled to the MSSM, leading to calculable soft supersymmetry breaking masses and a distinct low energy phenomenology that allows to differentiate it from other models of supersymmetry breaking and mediation mechanisms.
  \end{abstract}

  \section{Introduction}

  In a recent work~\cite{RK}, we studied a simple $N=1$ supergravity model of supersymmetry breaking~\cite{Z1} having a metastable de Sitter vacuum with an infinitesimally small (tunable) cosmological constant independent of the supersymmetry breaking scale that can be in the TeV region. Besides the gravity multiplet, the minimal field content consists of a chiral multiplet with a shift symmetry promoted to a gauged R-symmetry using a vector multiplet. In the string theory context, the chiral multiplet can be identified with the string dilaton (or an appropriate compactification modulus) and the shift symmetry associated to the gauge invariance of a two-index antisymmetric tensor that can be dualized to a (pseudo)scalar. The shift symmetry fixes the form of the superpotential and the gauging allows for the presence of a Fayet-Iliopoulos (FI) term, leading to a supergravity action with two independent parameters that can be tuned so that the scalar potential possesses a metastable de Sitter minimum with a tiny vacuum energy (essentially the relative strength between the F- and D-term contributions). A third parameter fixes the Vacuum Expectation Value (VEV) of the string dilaton at the desired (phenomenologically) weak coupling regime. An important consistency constraint  of our model is anomaly cancellation which has been studied in~\cite{R2} and implies the existence of additional charged fields under the gauged R-symmetry. 

In this work, we study a small variation of this model which is manifestly anomaly free without additional charged fields and allows to couple in a straight forward way a visible sector containing the minimal supersymmetric extension of the Standard Model (MSSM) and study the mediation of supersymmetry breaking and its phenomenological consequences. It turns out that an additional `hidden sector' field $z$ is needed to be added for the matter soft scalar masses to be non-tachyonic; although this field participates in the supersymmetry breaking and is similar to the so-called Polonyi field, it does not modify the main properties of the metastable de Sitter vacuum. All soft scalar masses, as well as trilinear A-terms, are generated at the tree level and are universal under the assumption that matter kinetic terms are independent of the `Polonyi' field, since matter fields are neutral under the shift symmetry and supersymmetry breaking is driven by a combination of the $U(1)$ D-term and the dilaton and $z$-field F-term. Alternatively, a way to avoid the tachyonic scalar masses without adding the extra field $z$ is to modify the matter kinetic terms by a dilaton dependent factor.

A main difference of the present analysis from the previous work is that we use a field representation in which the gauged shift symmetry corresponds to an ordinary $U(1)$ and not an R-symmetry. The two representations differ by a K\"ahler transformation that leaves the classical supergravity action invariant. However, at the quantum level, there is a Green-Schwarz term generated that amounts an extra dilaton dependent contribution to the gauge kinetic terms needed to cancel the anomalies of the R-symmetry. This creates an apparent puzzle with the gaugino masses that vanish in the first representation but not in the latter. The resolution to the puzzle is based to the so called anomaly mediation contributions~\cite{gauginomass,bagger} that explain precisely the above apparent discrepancy. It turns out that gaugino masses are generated at the quantum level and are thus suppressed compared to the scalar masses (and A-terms).

The outline of the paper is the following. In Section 2, we present the model and our conventions and show that adding MSSM fields inert under the shift symmetry leads to tachyonic scalar masses. In Section 3, we solve this problem by extending the model with an additional chiral field in the `hidden' sector, participating in the supersymmetry breaking without modifying the main features of the model and its metastable de Sitter vacuum. In Section 4, we add a visible sector with the MSSM fields and compute all soft breaking terms. In particular, we discuss how gaugino masses are generated and describe the puzzle mentioned above. In Section 5, we discuss the K\"ahler transformation and show the equivalence of the two representations at the quantum level. We work out the phenomenology in section 6. In section 7 we introduce a non-canonical K\"ahler potential for the MSSM superfields as a different possible solution to the tachyonic masses. Section 8 contains our conclusions. Appendix A contains the computation of the fermion mass matrix in the models of Sections 3 and 4, while Appendix B describes the anomaly cancellation.

  \section{Conventions} \label{sec:conventions}

  Throughout this paper we use the conventions of \cite{VP}. A supergravity theory is specified (up to Chern-Simons terms) by a K\"ahler potential $\mathcal K$, a superpotential $W$, and the gauge kinetic functions $f_{AB}(z)$. The chiral multiplets $z^\alpha, \chi^\alpha$ are enumerated by the index $\alpha$ and the indices $A,B$ indicate the different gauge groups. Classically, a supergravity theory is invariant under K\"ahler tranformations, viz.
  \begin{align} \mathcal K(z ,\bar z) &\longrightarrow \mathcal K(z , \bar z) + J(z) + \bar J(\bar z), \label{kahler_tranformation} \notag \\ W(z) &\longrightarrow e^{-\kappa^2 J(z)} W(z), \end{align}
  where $\kappa$ is the inverse of the reduced Planck mass, $m_p = \kappa^{-1} = 2.4 \times 10^{15} $ TeV. The gauge transformations of chiral multiplet scalars are given by holomorphic Killing vectors, i.e. $\delta z^\alpha = \theta^A k_A^\alpha (z)$, where $\theta^A$ is the gauge parameter of the gauge group $A$.   The K\"ahler potential and superpotential need not be invariant under this gauge transformation, but can change by a K\"ahler transformation
  \begin{align} \delta \mathcal K &= \theta^A \left[r_A(z) + \bar r_A(\bar z)\right], \ \end{align}
  provided that the gauge transformation of the superpotential satisfies $\delta W = - \theta^A \kappa^2 r_A(z) W $. One then has from $\delta W = W_\alpha \delta z^\alpha$ 
  \begin{align} W_\alpha k_A^\alpha = - \kappa^2 r_A W, \label{Wtransf} \end{align}
  where $W_\alpha = \partial_\alpha W$ and $\alpha$ labels the chiral multiplets.  
  The supergravity theory can then be described by a gauge invariant function
  \begin{align} \mathcal G = \kappa^2 \mathcal K + \log(\kappa^6 W \bar W). \end{align}

  The scalar potential is given by
  \begin{align} V &= V_F + V_D \notag \\ V_F &= e^{\kappa^2 \mathcal K} \left( - 3 \kappa^2 W \bar W  + \nabla_\alpha W g^{\alpha \bar \beta} \bar \nabla_{\bar \beta} \bar W \right)  \notag \\ V_D &= \frac{1}{2} \left( \re f \right)^{-1 \ AB} \mathcal P_A \mathcal P_B,\end{align}
  where W appears with its K\"ahler covariant derivative 
  \begin{align}\nabla_\alpha W = \partial_\alpha W(z) + \kappa^2 (\partial_\alpha \mathcal K) W(z).\end{align}
  The moment maps $\mathcal P_A$ are given by
  \begin{align} \mathcal P_A = i(k_A^\alpha \partial_\alpha \mathcal K - r_A). \label{momentmap} \end{align}
  In this paper we will be concerned with theories having a gauged R-symmetry, for which $r_A(z)$ is given by an imaginary constant $r_A(z) = i \kappa^{-2} \xi $. In this case, $\kappa^{-2} \xi$ is a Fayet-Iliopoulos \cite{FI} constant parameter.

  \section{Introduction of the model}\label{sec:motivation}
  \subsection{Motivation}
  In \cite{RK,Z1}  a class $\mathcal N = 1$ supergravity theories based on a gauged R-symmetry which allow for metastable de Sitter (dS) vacua was presented. These theories have a tunable (infintesimally small) value of the cosmological constant and a TeV gravitino mass.
  The spectrum consists, in addition to the supergravity multiplet, of a chiral multiplet $S$ and a vector multiplet associated with a shift symmetry of the scalar component $s$ of the chiral multiplet $S$
  \begin{align} \delta s = -ic \theta. \label{shift} \end{align}
  
  The goal of this paper is to generalize this model such that it is anomaly-free and can be coupled to the MSSM and make phenomenological predictions, while maintaining its desirable properties described in \cite{RK,Z1} such as a tunable cosmological constant and a TeV gravitino mass.
  
  The starting point is a chiral multiplet $S$ invariant under a gauged shift symmetry (\ref{shift}) and a string-inspired K\"ahler potential of the form $-p\log(s+ \bar s)$. The most general superpotential\footnote{This was already noticed in \cite{Gomez}.} is either a constant $W=\kappa^{-3}a$ or an exponential superpotential $W=\kappa^{-3}ae^{bs}$ (where $a$ and $b$ are constants). A constant superpotential is (obviously) invariant under the shift symmetry, while an exponential superpotential transforms as $W \rightarrow W e^{-ibc\theta}$, as in eq. (\ref{Wtransf}).   In this case the shift symmetry becomes a gauged R-symmetry and the scalar potential contains a Fayet-Iliopoulos term.   Note however that by performing a K\"ahler transformation (\ref{kahler_tranformation}) with $J=\kappa^{-2}bs$, the model can be recast into a constant superpotential at the cost of introducing a linear term in the K\"ahler potential $\delta K=b(s+\bar s)$.   Even though in this representation, the shift symmetry is not an R-symmetry, we will still refer to it as $U(1)_R$.   
  The most general gauge kinetic function has a constant term and a term linear in $s$, $f(s)=\delta + \beta s$. 
 
 To summarise,\footnote{  In superfields the shift symmetry (\ref{shift}) is given by $\delta S = -ic \Lambda$, where $\Lambda$ is the superfield generalization of the gauge parameter.   The gauge invariant K\"ahler potential is then given by $\mathcal K(S , \bar S) = -p \kappa^{-2} \log (S+\bar S+cV_R)+\kappa^{-2}b(S+\bar S+c V_R)$, where $V_R$ is the gauge superfield of the shift symmetry.   }
  \begin{align}    \mathcal K(s,\bar s) &= -p \log(s + \bar s) + b(s + \bar s), \notag \\   W(s) &= \kappa^{-3} a,\notag \\   f(s) &= \delta + \beta s\, ,  \end{align}
where the constants $a$ and $b$ together with the constant $c$ in eq. (\ref{shift}) can be tuned to allow for an infinitesimally small cosmological constant and a TeV gravitino mass. For $b>0$, there always exists a supersymmetric AdS (anti-de Sitter) vacuum at $\langle s + \bar s \rangle = b/p$. We therefore focus on $b<0$.   In the context of string theory, $S$ can be identified with a compactification modulus or the universal dilaton and (for negative $b$) the exponential superpotential may be generated by non-perturbative effects.

For $p \geq 3$ the scalar potential V is positive and monotonically decreasing \cite{Z1}, while for $p<3$, its F-term part $V_F$ is unbounded from below when $s + \bar s \rightarrow 0$. On the other hand, the D-term part of the scalar potential $V_D$ is positive and diverges when $s + \bar s \rightarrow 0$ and for various values for the parameters an (infinitesimally small) positive (local) minimum of the potential can be found.

If we restrict ourselves to integer $p$, tunability of the vacuum energy restricts $p=2$ or $p=1$ when $f(s)=s$, or $p=1$ when the gauge kinetic function is constant. 

Let us first consider $\beta \neq 0$: The case when $p=2$ and $f(s)=s$ has been analyzed in full detail in \cite{RK}. For a field-dependent gauge kinetic function, the Lagrangian contains a Green-Schwarz \cite{Green-Schwarz} term
  \begin{align}  \mathcal L_{GS} &=  \frac{1}{8} \text{Im}(f(s)) \epsilon^{\mu \nu \rho \sigma} F_{\mu \nu} F_{\rho \sigma}, \label{LGS1} \end{align} 
Since this term is not invariant under the shift symmetry (\ref{shift}),
\begin{align} \delta \mathcal L_{GS} =  -\theta \frac{\beta c}{8}   \ \epsilon^{\mu \nu \rho \sigma} F_{\mu \nu} F_{\rho \sigma}. \end{align}
 its variation should be canceled. As explained in Appendix \ref{Appendix:cubic}, in the 'frame' with an exponential superpotential the R-charges of the fermions in the model can give an anomalous contribution to the Lagrangian. In this case the Green-Schwarz term can cancel quantum anomalies. However as shown in \cite{R2}, with the minimal MSSM spectrum, the presence of the term (\ref{LGS1}) requires the existence of additional fields in the theory charged under the shift symmetry.

Instead, to avoid the discussion of anomalies at this point, we focus on models with a constant gauge kinetic function. In this case the only (integer) possibility\footnote{ If $f(s)$ is constant, the leading contribution to $V_D$ when $s + \bar s \rightarrow 0$ is proportional to $1/(s + \bar s)^2$, while the leading contribution to $V_F$ is proportional to $1/(s + \bar s)^p$. It follows that $p<2$; if $p>2$, the potential is unbounded from below, while if $p=2$, the potential is either positive and monotonically decreasing or unbounded from below when $s+ \bar s \rightarrow 0$ depending on the values of the parameters.} is $p=1$. However, as we will show below, this model suffers from tachyonic soft masses when it is coupled to the MSSM. 

\subsection{Models with field-independent gauge kinetic functions} \label{sec:model}
    
As described above, 
a constant gauge kinetic function dictates $p=1$. Moreover, by appropriate field redefinitions, this constant can be absorbed in the other constants of the theory. We can therefore take $f(s)=1$.  As also described above, 
the model with an exponential superpotential can be recast by a K\"ahler transformation in a model with a constant superpotential, but with a linear term in the K\"ahler potential. To avoid any quantum anomalies coming fron the R-charges of the various fermions in the model, we continue with a constant superpotential and a linear term in $s+\bar s$ in the K\"ahler potential.  Although these models are equivalent classically, they might differ at the quantum level. The model is given by  
\begin{align} \mathcal K &= - \kappa^{-2} \log (s + \bar s) + \kappa^{-2} b (s + \bar s), \notag \\ W &= \kappa^{-3} a , \notag \\  f(s) &= 1\, . \label{model0} \end{align}  
The scalar potential is given by
  \begin{align} V &= V_F + V_D, \notag \\ V_F &= \kappa^{-4} |a|^2 \frac{e^{b(s + \bar s)}}{s + \bar s} \sigma_s, && \sigma_s=  -3 + \left( b(s + \bar s) - 1 \right)^2 , \notag \\ V_D &= \kappa^{-4}   \frac{c^2}{2} \left( b - \frac{1}{s + \bar s} \right)^2. \label{scalarpot0}\end{align}

  As mentioned in the previous subsection, 
for $b>0$ this scalar potential always allows for a supersymmetric AdS minimum at $\langle s + \bar s \rangle = 1/b$, while for $b=0$ supersymmetry is broken in AdS space~\cite{RK}. We therefore focus on the case $b<0$. The minimization of the potential $\partial_s V = 0$ gives
  \begin{align} \frac{c^2}{a^2} &=  \langle s + \bar s \rangle (2- b^2 \langle s + \bar s\rangle ^2) e^{b \langle s + \bar s \rangle}\, . \end{align}
  By plugging this relation into $V_{min} = \Lambda \approx (10^{-3} \text{eV})^4$, one finds
  \begin{align} \kappa^4 e^{-b \langle s + \bar s \rangle} \langle s + \bar s \rangle \frac{\Lambda}{a^2} = -3 + (b \langle s + \bar s  \rangle -1 )^2 \left[ 2 - \frac{b^2 \langle s + \bar s \rangle^2}{2}\right]. \end{align}
  An infinitesimally small cosmological constant $\Lambda$ can then be obtained by tuning the parameters $a,b,c$ such that
  \begin{align} b \langle s + \bar s \rangle &= \alpha \approx -0.233153, \notag \\ \frac{b c^2}{a^2} &= A(\alpha) + \frac{2 \kappa^4 \Lambda \alpha^2}{a^2 b (\alpha - 1)^2}, & A(\alpha) &=  2 e^\alpha \alpha \frac{ 3 - (\alpha -1)^2}{ (\alpha - 1)^2} \approx -0.359291\, , \label{bsalpha} \end{align}
  where $\alpha$ is the negative root of $-3 + (\alpha -1 )^2(2 - \alpha^2/2) = 0$ close to $-0.23$. The other roots are either imaginary or would not allow for a real solution of the second constraint.

  We conclude that this model allows for a stable de Sitter (dS) vacuum with an infinitesimally small (and tunable) value for the cosmological constant.

  Unfortunately, if one now adds an MSSM-like field $\varphi$ with a canonical K\"ahler potential, vanishing superpotential and invariant under the shift symmetry of the model,
 \begin{align} \mathcal K &= - \kappa^{-2} \log (s + \bar s) + \kappa^{-2} b (s + \bar s) + \sum \varphi \bar \varphi, \notag \\ W &= \kappa^{-3} a + W_{MSSM} ,  \label{model0MSSM} \end{align}  
where 
$W_{MSSM}$ is the MSSM superpotential defined below in eq. (\ref{MSSMsuperpot}),
 the soft scalar mass squared at $\langle \varphi \rangle = \langle \bar \varphi \rangle = 0$ is negative, given by
  \begin{align}  \left. \partial_\varphi \partial_{\bar \varphi} V \right|_{\langle \varphi \rangle = 0}=   |a|^2 b \frac{e^{\alpha}}{\alpha} \left(\langle \sigma_s \rangle +1 \right) < 0\ . \label{tach} \end{align}
  Since $\langle \sigma_s \rangle \approx -1.48$, any nonzero solutions $\langle \varphi \rangle \neq 0$ of $\partial_\varphi V = 0$ would mean that the field $\varphi$ contributes in general to the supersymmetry breaking. We conclude that the model on its own can not be consistently extended to include the MSSM with canonical kinetic terms.  To circumvent this problem, one can add an extra hidden sector field which contributes to (F-term) supersymmetry breaking. This will be worked out in full detail in the following sections.  However, we will show in section \ref{sec:noncan} that the problem of tachyonic soft masses can also be solved if one allows for a non-canonical K\"ahler potential in the visible sector, which gives an additional contribution to the masses through the D-term.   
 
\section{Extra field in the hidden sector} \label{sec:extrahidden}
  \subsection{Tuning of the parameters}
  As described above, the model (with $p=1$ and a field independent gauge kinetic function) presented there would give a tachyonic mass to any MSSM-like fields (that are invariant under the shift symmetry and have a canonical K\"ahler potential).  In this section we add an extra hidden sector field $z$ (similar to the so-called Polonyi field \cite{polonyi}) to circumvent this problem. Note that this choice is not unique and that the problem can also be circumvented by allowing a non-canonical K\"ahler potential for the MSSM fields (see section \ref{sec:noncan}).

 The K\"ahler potential, superpotential and gauge kinetic function are given by
  \begin{align} \mathcal K &= -\kappa^{-2} \log (s + \bar s) + \kappa^{-2} b (s + \bar s) + z \bar z, \notag \\ W &= \kappa^{-3} a (1+ \gamma \kappa z) , \notag \\ f(s) &= 1\, , \label{model} \end{align}
  with $\gamma$ an additional constant parameter.  The scalar potential is
  \begin{align} V &= V_F + V_D, \notag \\ V_F &= \kappa^{-4} |a|^2 \frac{e^{b(s + \bar s) + \kappa^2 z \bar z}}{s + \bar s} \left( \sigma_s A(z,\bar z) + B(z, \bar z) \right),  \notag \\ V_D &= \kappa^{-4}   \frac{c^2}{2} \left( b - \frac{1}{s + \bar s} \right)^2, \end{align}
  where
  \begin{align}  A(z, \bar z) &= \left| 1 + \gamma \kappa z \right|^2, \notag \\  B(z, \bar z) &= \left| \gamma + \kappa \bar z + \gamma \kappa^2 z \bar z \right|^2.  \end{align}
  We focus on real $z=\bar z = \kappa^{-1} t$: 
  \begin{align} A(t) &= (1+\gamma t)^2, \notag \\ B(t) &= (\gamma + t + \gamma t^2)^2\, ; \label{AB} \end{align} 
  $\partial_t V = 0$ then gives
  \begin{align} 0 &= \gamma(\sigma_s\! +\!1) + (\sigma_s\! +\! 1\! +\! \gamma^2 (\sigma_s\! +\! 2) ) t + \gamma (2 \sigma_s +5) t^2 + ( 1 + \gamma^2 (\sigma_s\! +\! 4) ) t^3\! +\! 2 \gamma t^4\! +\! \gamma^2, \notag \\ \sigma_s &= -3 + (\alpha - 1)^2,  \ \ \ \ \ \ \ \ \  \alpha = b(s + \bar s). \label{Vt} \end{align} 
  As in the previous section, $\partial_s V = 0$, implies
  \begin{align}  \frac{c^2}{a^2 } =\frac{\alpha}{b} e^{\alpha + t^2} \left[ A( t ) ( 2 - \alpha^2 ) - B(t) \right]. \label{ca} \end{align}
  This can be combined with $V = 0$
  \begin{align}  \frac{c^2}{a^2} = -2 \frac{\alpha}{b} e^{\alpha + t^2} \left[\frac{\sigma_s A(t) + B(t)}{(\alpha-1)^2} \right], \label{ca2} \end{align}
  to give
  \begin{align} 0 &= A(t) \left( \sigma_s - \frac{1}{2} (\alpha - 1)^2 (\alpha-2) \right) + B(t) \left(1 - \frac{1}{2} (\alpha - 1)^2 \right). \label{Vs} \end{align}
  In princple for any value of $\gamma$, a Minkowski minimum can be found by solving eqs. (\ref{Vt}) and (\ref{Vs}) for $\alpha$ and $t$, and then tuning the parameters $a$,$b$ and $c$ by using the relation (\ref{ca2}). 

  The role of the extra hidden sector field $z$ is to give a (positive) F-term contribution to the  scalar potential, which in turn gives a positive contribution (proportional to $\left| \nabla_z W\right|^2$) to the soft mass squared of any MSSM-like field in eq. (\ref{tach}). It turns out that the addition of the extra hidden sector field $z$ indeed results in positive soft masses squared. 

  It is however necessary that $z$ contributes to the supersymmetry breaking. The existence of any minimum of the potential with $\left| \nabla_z W \right|^2 =0$ can be troublesome and we therefore require
  \begin{align} \nabla_z W = \partial_z W + \kappa^2 \mathcal K_z W = a\left(\gamma + \bar z (1+ \gamma z) \right) \neq 0 \label{noAdS} \end{align}
  Since $\gamma$ is real, any root of $\nabla_z W =0$ is also real. To ensure the condition (\ref{noAdS}) we must ensure that the roots $ \text{Re}(z) =( -1 \pm \sqrt{1-4 \gamma})/4\gamma$ are complex. This requires $|\gamma| > 1/2$.

  Also, for any $\gamma$ the solution $(\alpha, t)$ of the set of equations  (\ref{Vt}) and (\ref{Vs}) should give a positive right hand side of eq. (\ref{ca}) (or equivalently, eq. (\ref{ca2})). This constraint leads to $\gamma < 1.707$. We conclude that
  \begin{align} \gamma \in \left[  0.5 ,1.707 \right]. \label{gammarange} \end{align} 

  For example, for $\gamma = 1$, we have $b \langle s + \bar s \rangle = \alpha \approx -0.134014$, $\langle t \rangle = 0.39041$. The (negative) constant $b$ can be chosen freely to fix the value of the vacuum expectation value (VEV) of Re$(s)$. The parameters $a$ and $c$ should be tuned carefully according to
  \begin{align} \frac{b c^2}{a^2} = -2 \alpha e^{\alpha + t^2} \left[\frac{\sigma_s A(t) + B(t)}{(\alpha-1)^2} \right]	\approx -0.1981	. \label{ca1} \end{align}
  Note that the number on the right hand side changes when $\gamma$ is varied. The remaining free parameter $a$ can be used to tune the supersymmetry breaking scale and (as shown below) the soft masses for the MSSM-like fields compared to the gravitino mass depend slightly on $\gamma$  (provided $c$ and $a$ are also tuned according to eq. (\ref{ca})). We summarise the VEVs of $\alpha$ and $t$, together with the above constraint on the parameters for the particular  choice $\gamma=1$ below for future reference
  \begin{align}
   \gamma = 1, \ \ \alpha \approx -0.134014, \ \ \langle t \rangle \approx 0.39041, \ \ \frac{bc^2}{a^2} \approx -0.1981 \ . \label{parameterchoice}
  \end{align} 

  For $\gamma$ in the allowed parameter range (\ref{gammarange}), the scalar potential is positive definite for all Re$(s)>0, z, \bar z$, including the imaginary part of $z$, which justifies our assumption to look for a Minkowski minimum with Im$(z) = 0$. 
  In fact, for the allowed values of $\gamma$, the solution of the set of equations (\ref{Vt}) and (\ref{Vs}) together with $\partial_{\text{Im}(z)}V=0$ gives $\text{Im}(z)=0$ as a solution. 

  Finally, note that this Minkowski minimum can be lifted to a dS vacuum with an infinitesimally small cosmological constant by a small increase in $c$. A cosmological constant $\Lambda$ can be obtained by replacing the condition (\ref{ca1}) with
  \begin{align} \frac{c^2}{a^2 } =-2 \frac{\alpha}{b} e^{\alpha + t^2} \left[\frac{\sigma_s A(t) + B(t)}{(\alpha-1)^2} \right] + \frac{2 \alpha^2}{(\alpha-1)^2} \frac{\kappa^4 \Lambda}{a^2 b^2} . \end{align}

  \subsection{Scalar masses, gravitino mass, super-BEH and St\"uckelberg mechanism.}

  The gravitino mass is given by
  \begin{align} m_{3/2} = \kappa^2 e^{\kappa^2 \mathcal K/2} W =  \kappa^{-1} a \sqrt { \frac{b}{ \alpha} } e^{\alpha/2 + t^2/2} \left( 1 + \gamma t \right).  \label{gravitino} \end{align}
  Note that this can be arranged to be at the TeV scale by suitably tuning $a$. For example, for $\gamma=1$, such that $\alpha$ and $t$ are given by eq. (\ref{parameterchoice}) and $m_{3/2} =1 \text{ TeV}$, we have
  \begin{align}
 a \sqrt b \approx 3.53 \times 10^{-17}. \label{a_numerics}   
  \end{align}
Since the VEV of Im$(z)$ vanishes, it does not mix with the other hidden sector scalars and its mass is given by
\begin{align}
 m^2_{\text{Im}(z)} &= m_{3/2}^2 \  f_{\text{Im}(z)}, \notag \\
   f_{\text{Im}(z)} &=\frac{2  \left(1+2 t^3 \gamma +t^4 \gamma ^2+\sigma_s +2 t \gamma  (2+\sigma_s )+\gamma ^2 (3+\sigma_s )+t^2 \left(1+\gamma ^2 (4+\sigma_s )\right)\right)}{(1 + \gamma t)^2 }.
\end{align}
However, the masses of the scalars Re$(s)$ and Re$(z)$ mix, so one should diagonalize their mass matrix (with eigenvalues $m_{ts1}$ and $m_{ts2}$) while taking in account the non-canonical kinetic term for $s$.  We omit the details and merely state the result for the particular choice of parameters $\gamma = 1$ in eq. (\ref{parameterchoice}):
  \begin{align}   m_{\text{Im}(z)} \approx 1.21\ m_{3/2}, \notag \\   m_{ts1} \approx 4.34\ m_{3/2}, \notag \\  m_{ts2} \approx 1.08\ m_{3/2} . \label{scalarmasses}   \end{align}

  The imaginary part of $s$ is eaten by the $U(1)$ gauge boson, which becomes massive. Its mass is given by\footnote{ This is calculated as follows: The relevant part of the Lagrangian is   \begin{align} \mathcal L/e = -\frac{1}{(s + \bar s)^2} \left(\partial_\mu s + ic A_\mu \right) \left( \partial^\mu s - ic A^\mu \right) - \frac{f(s)}{4} F_{\mu \nu} F^{\mu \nu}. \notag \end{align}   Use the gauged shift symmetry to put $\text{Im}(s) = 0$ and obtain   \begin{align} \mathcal L/e = -\frac{1}{(s + \bar s)^2} \partial_\mu \text{Re}(s) \partial^\mu \text{Re}(s) - \frac{1}{4}  F_{\mu \nu} F^{\mu \nu} - \frac{c^2}{(s + \bar s)^2} A_\mu A^\mu. \notag \end{align} }:
 \begin{align} m_{A_\mu} &= \frac{\kappa^{-1} b c }{\alpha} \notag \\ &\approx 0.87 \ m_{3/2},\label{gaugebosonmass} \end{align} 
where the last line was obtained by the relation between the parameters eq. (\ref{ca1}) and by substituting the numerical values for $\gamma=1$ eq. (\ref{parameterchoice}).

  The Goldstino, which is a linear combination of the gaugino, the z-fermion and the s-fermion, is eaten by the gravitino, which in turn becomes massive. The masses of the remaining two hidden sector fermions are calculated in Appendix \ref{Appendix:Fermions} and their values for $\gamma=1$ are given by
\begin{align}  m_{\chi_1} & \approx   2.27 \ m_{3/2}, \notag \\  m_{\chi_2} & \approx  0.12 \ m_{3/2}. \end{align}
  
  \subsection{Tree level soft masses}

  The goal of this section is to use the coupling of the model above, that allows for a TeV gravitino and an infinitesimally small cosmological constant, to the MSSM and to calculate its soft breaking terms.

As already said, for simplicity, we take the MSSM-like fields $\varphi_\alpha$ to be chargeless under the extra $U(1)$. They can then easily be coupled to the above model in the following way:
  \begin{align} \mathcal K &= - \kappa^{-2} \log (s + \bar s) + \kappa^{-2} b (s + \bar s) + z \bar z + \sum_\alpha \varphi \bar \varphi , \notag \\ W &= \kappa^{-3} a (1+\kappa z) + W_{\text{MSSM}}(\varphi) , \notag \\ f_R(s) &= 1, \ \  \ \ \ \ \ \ \ \ \ \  \ f_A(s) = 1/g_A^2. \label{modelMSSM} \end{align}
  The various multiplets in the MSSM are labeled by an (omitted for simplicity) index $\alpha$. The Standard Model gauge groups are labeled by an index A, while the extra $U(1)$ will be referred to with an index $R$. Note that all gauge kinetic functions are taken to be constants. 

  The scalar potential is now given by
      \begin{align} V &= V_F + V_D, \notag \\ V_F &= \kappa^{-4} \frac{e^{b(s + \bar s) + z\bar z + \varphi \bar \varphi} }{s + \bar s} \left( \sigma_s A(z,\bar z, \varphi, \bar \varphi) + B(z, \bar z, \varphi, \bar \varphi) + \kappa^4 \sum_\alpha \left| \nabla_\alpha W \right|^2 \right),  \notag \\ V_D &=  \kappa^{-4} \frac{c^2}{2} \left( b - \frac{1}{s + \bar s} \right)^2, \label{scalarpotMSSM} \end{align}
      where
      \begin{align}   A(z,\bar z, \varphi, \bar \varphi) &= \left| a + a \gamma \kappa z + \kappa^3 W_{\text{MSSM} } \right|^2 \notag \\  B(z, \bar z, \varphi, \bar \varphi) &= \left| a \gamma + \kappa \bar z (a + a \gamma z + \kappa^3 W_{\text{MSSM} } ) \right| ^2 \notag \\   \left| \nabla_\alpha W \right|^2 &= \left| \partial_\alpha W_{\text{MSSM}} + \kappa^2 \bar \varphi W \right|^2\, .  \end{align}
  It can be easily seen that the resulting scalar potential has a minimum at $\langle \varphi \rangle = \langle W_{\text{MSSM}} \rangle =  0$, in which case the potential of last section is reproduced and its conclusions are still valid.   For example, $A(z,\bar z, \varphi,\bar \varphi)|_{\langle z \rangle = t, \langle \varphi \rangle = 0} = a^2 A(t)$ and $B(z,\bar z, \varphi,\bar \varphi)|_{\langle z \rangle = t, \langle \varphi \rangle = 0} = a^2 B(t)$, where $A(t)$ and  $B(t)$ are defined in eq. (\ref{AB}).   The second derivatives of the potential, evaluated on the ground state are given by 
  \begin{align}   \partial_\varphi \partial_{\bar \varphi} V &= \frac{\kappa^{-2} a^2 b e^{\alpha + t^2}}{\alpha} \left[ (\sigma_s + 1 ) A(t) + B(t)  + \kappa^2 W_{\varphi \varphi} \bar W_{\bar \varphi \bar \varphi} \right], \notag \\  \partial_\varphi \partial_\varphi V &= \frac{ \kappa^{-1} ab W_{\varphi \varphi} e^{\alpha + t^2}}{\alpha}   \left[ (\sigma_s +2) (1+ \gamma t) + t (\gamma +t + \gamma t^2)  \right].  \end{align}
  There is no mass mixing between the different $\varphi^\alpha$ (except of course for the $B_0$ term defined below) and between the MSSM fields with $z$ and $s$. Let us now specify the MSSM superpotential
  \begin{align} W_{MSSM}= y_u^{ij} \bar u_i Q_j \cdot H_u - y_d^{ij} \bar d_i Q_j \cdot H_d - y_e^{ij} \bar e_i L_j \cdot H_d + \mu H_u \cdot H_d. \label{MSSMsuperpot} \end{align}
  Note that in the scalar potential eq. (\ref{scalarpotMSSM}) the MSSM F-terms $\sum_\alpha \left| \nabla_\alpha W \right|^2 $ come with a prefactor $\exp (\alpha  + t^2) b/\alpha$ (where the fields have been replaced by their VEVs). To bring this into a conventional form, one should rescale the MSSM superpotential
  \begin{align}  \hat W_{MSSM} = \sqrt{\frac{b} {\alpha}}  e^{\alpha/2 + t^2/2 } \ W_{MSSM}.  \label{MSSMhat}  \end{align}
  Then the squark and slepton soft masses are given by
  \begin{align}  m^2_{\tilde Q} &= m^2_{\tilde {\bar u}}= m^2_{\tilde {\bar d}}= m^2_{\tilde Q} = m^2_{\tilde {\bar Q}} = m_0^2 \ \mathbb{I}, \notag \\  m_0^2 &= \kappa^{-2} b a^2 \frac{e^{\alpha + t^2 } }{\alpha} \left[ A(t) \left( \sigma_s + 1 \right) + B(t)\right].  \end{align}
  Here, $\mathbb{I}$ is the unit matrix in family space.  The trilinear couplings are given by 
  \begin{align}  &a_u = A_0 \hat y_u, \ \ \ \ \ \ \  a_d = A_0  \hat y_d, \ \ \ \ \ \ \ a_e = A_0 \hat y_e, \notag \\  &A_0= \kappa^{-1}a \sqrt{\frac{b}{\alpha}} e^{(\alpha + t^2)/2 }  \left[  (\sigma_s +3) (1+\gamma t) + t(\gamma + t +\gamma t^2) \right],  \end{align}
  where $\hat y_u, \hat y_d $ and $ \hat y_e$ are the Yukawa couplings of the MSSM superpotential after the rescaling of eq. (\ref{MSSMhat}).  Also,
  \begin{align} &m^2_{H_u } = m^2_{H_d} = m_0^2. \end{align}
  and 
  \begin{align}  &B_0 = \kappa^{-1}a \sqrt{\frac{b}{\alpha}} e^{(\alpha  + t^2)/2}  \left[  (\sigma_s + 2)(1+\gamma t) + t (\gamma + t + \gamma t^2)  \right],  \end{align}
  where $B_0$ generates a term proportional to $- \hat \mu B_0  H_u \cdot H_d + \text{h.c.}$, where $\hat \mu$ is the rescaled $\mu$-parameter (in the sense of eq. (\ref{MSSMhat})).  Summarised, in terms of the gravitino mass (eq. (\ref{gravitino})), the MSSM soft terms are given by
  \begin{align}
  m_0^2 &= m_{3/2}^2  \left[ \left( \sigma_s + 1 \right) + \frac{(\gamma + t + \gamma t)^2}{(1 + \gamma t)^2}\right], \notag \\   A_0&= m_{3/2} \left[  (\sigma_s +3)  + t \frac{ (\gamma + t +\gamma t^2) } {1+\gamma t} \right], \notag \\   B_0 &= m_{3/2}  \left[  (\sigma_s + 2) + t \frac{ (\gamma + t + \gamma t^2) }{(1+\gamma t)}  \right]. \label{softterms}   \end{align}
  Note the relation \cite{A-B relation}
  \begin{align}   A_0 = B_0 + m_{3/2}. \label{AB-relation}   \end{align}
  At tree level, the gaugino masses are given by
  \begin{align}
  m_{AB} = -\frac{1}{2} e^{\kappa^2 \mathcal K/2} f_{AB,\alpha} g^{\alpha \bar \beta} \bar \nabla_{\bar \beta} \bar W\, , \label{mgaugino}
  \end{align}
where the indices $A$ and $B$ label the different gauge groups and $f_{AB,\alpha}$ stands for $\partial_\alpha f_{AB}$. Since the gauge kinetic functions are constant, they vanish
  \begin{align} \left. m_{AB} \right|_{\text{tree}} = 0. \end{align}

 However, as mentioned in section \ref{sec:motivation}, the K\"ahler potential and superpotential of any ($\mathcal N = 1, D= 4$) supergravity theory are only determined up to K\"ahler transformations, at least classically.\footnote{This statement is only true for supergravity theories with a non-vanishing superpotential where everything can be defined in terms of a gauge invariant function $G = \kappa^2 \mathcal K + \log(\kappa^6 W \bar W)$ \cite{W=0}.} By applying a K\"ahler transformation (\ref{kahler_tranformation}) with $J= -\kappa^{-2} b s$ to the model defined in eq. (\ref{modelMSSM}), one ends up with the classically equivalent theory
  \begin{align}  \mathcal K &=- \kappa^{-2} \log (s + \bar s) + z \bar z + \sum_\alpha \varphi \bar \varphi   , \notag \\   W &=\left( \kappa^{-3} a (1+z) + W_{\text{MSSM}}(\varphi) \right) e^{bs}. \label{KahlertransformedMSSM}   \end{align}
  Note that all classical results of the previous section also hold for this theory: Its scalar potential is given by (\ref{scalarpotMSSM}) and can be tuned in exactly the same way as above.   In particular, the $A_0, B_0$ and $m_0$ soft terms are again given by eqs. (\ref{softterms}).   However, since a K\"ahler transformation is anomalous \cite{KL}, there are in general additional contributions to the effective action at the quantum level. First note that the shift symmetry (\ref{shift}) of $s$ renders the superpotential non-gauge invariant
  \begin{align} W \longrightarrow W e^{-ibc\theta}. \end{align}
  In other words, the shift symmetry has become a gauged R-symmetry.  Therefore, all the fermions (including the gauginos and the gravitino) in the theory transform\footnote{The chiral fermions, the gauginos and the gravitino carry a charge $bc/2$, $-bc/2$ and $-bc/2$ respectively.} as well under this $U(1)_R$.   This leads to cubic $U(1)_R^3$ as well as mixed $U(1) \times G_{\text{MSSM}}$ anomalies.
 
 Anomalies in supergravity theories involving a gauged R-symmetry were carefully studied in \cite{R2,FreedmanAnomalies}; we summarise the main results in the Appendix \ref{Appendix:Anomalies}, where it has been shown that these anomalies are cancelled by a Green-Schwarz (GS) counter term. The latter arises from a quantum correction to the gauge kinetic functions given by\footnote{   Similarly, to cancel the cubic anomaly one should modify the R-gauge kinetic term as well to be $f_R(s) = 1 + \beta_R s$. It has been checked in Appendix \ref{Appendix:Anomalies} that $\beta_R = - \frac{35b^3 c^2}{96 \pi^2}$ is extremely small by eqs (\ref{ca1}) and (\ref{a_numerics}), so that the classical scalar potential (\ref{scalarpotMSSM}) is still valid to a very good approximation.  }
  \begin{align}  f_A(s) = 1/g_A^2 + \beta_A s. \label{gaugekineticfunction:fielddependent} \end{align}
  These field-dependent gauge kinetic functions give Green-Schwarz contributions
  \begin{align}  \mathcal L_{GS} &=  \frac{1}{8} \text{Im}(f_A(s)) \epsilon^{\mu \nu \rho \sigma} F_{\mu \nu}^A F_{\rho \sigma}^A, \notag \\
  \delta  \mathcal L_{GS} &= - \frac{\theta \beta_A c}{8}  \epsilon^{\mu \nu \rho \sigma} F_{\mu \nu}^A F^A_{\rho \sigma}.
  \end{align} 
  Anomaly cancellation then requires that (see eq. \ref{anomalieformules})
  \begin{align}   \beta_1 &= - \frac{  11 b}{8 \pi^2 }, \notag \\    \beta_2 &= - \frac{ 5 b }{8 \pi^2 }, \notag \\   \beta_3 &= - \frac{3 b}{8 \pi^2 }.    \end{align}
  The resulting gaugino masses are given by 
  \begin{align}   \hat M_1 &= \frac{11}{16\pi^2} b g_Y^2 e^{\alpha/2} (\alpha-1), \notag \\     \hat M_2 &= \frac{5}{16\pi^2} b g_2^2 e^{\alpha/2} (\alpha-1),\notag \\    \hat M_3 &= \frac{3}{16\pi^2} b g_3^2 e^{\alpha/2} (\alpha-1).\label{gauginomassdifference}   \end{align}
It is curious that the gaugino masses vanish for the model (\ref{modelMSSM}), while the classically equivalent model (\ref{KahlertransformedMSSM}) obtained upon a K\"ahler transformation has nonzero gaugino masses. This creates a puzzle on the quantum equivalence of these models. The answer to this puzzle is based on the fact that gaugino masses are present in both representations and are generated at one-loop level by an effect called Anomaly Mediation \cite{gauginomass,bagger}.  Indeed, it has been argued that gaugino masses receive a one-loop contribution due to the super-Weyl-K\"ahler and sigma-model anomalies. These contributions are different for both models, and we will show in section \ref{sec:modelR} that the difference accounts exactly for the contributions (\ref{gauginomassdifference}). Below, we compute the gaugino masses in the model (\ref{modelMSSM}) coming entirely from anomaly mediation.

The 'Anomaly Mediated' gaugino mass  contribution $M_{1/2}$ is given by \cite{bagger}
\begin{align}
  M_{1/2} = - \frac{g^2}{16 \pi^2} \left[ (3 T_G - T_R) m_{3/2} + (T_G - T_R) \mathcal K_\alpha F^\alpha + 2 \frac{T_R}{d_R} (\log \det \mathcal K|_R \ '')_{,\alpha} F^\alpha \right] , \label{gaugino mass}
  \end{align}
  where $T_G$ is the Dynkin index of the adjoint representation, normalized to $N$ for $SU(N)$, and $T_R$ is the Dynkin index associated with the representation $R$ of dimension $d_R$,   equal to $1/2$ for the $SU(N)$ fundamental. An implicit sum over all matter representations is understood. The quantity $3T_G - T_R$ is the one-loop beta function coefficient.   The expectation value of the auxiliary field $F^\alpha$, evaluated in the Einstein frame is given by   
  \begin{align}   F^\alpha = - e^{ \kappa^2 \mathcal K/2} g^{\alpha \bar \beta} \bar \nabla_{\bar \beta} \bar W.   \end{align}
  Clearly, for the K\"ahler potential (\ref{modelMSSM}) the last term in eq. (\ref{gaugino mass}) vanishes. However, the second term survives due to the presence of Planck scale VEVs for the hidden sector fields $s$ and $z$. By using the gravitino mass (\ref{gravitino}), the above expression can be rewritten as
 \begin{align}   M_{1/2} = - \frac{g^2}{16 \pi^2} m_{3/2} \left[(3 T_G - T_R) - (T_G - T_R) \left( (\alpha-1)^2 + t \frac{\gamma + t + \gamma t^2}{1 + \gamma t} \right) \right]  \end{align}
For $U(1)_Y$ we have $T_G = 0$ and $T_R = 11$, for $SU(2)$ we have $T_G = 2$ and $T_R = 7$, and for $SU(3)$ we have $T_G = 3$ and $T_R = 6$, such that for the different gaugino mass parameters this gives (in a self-explanatory notation):
 \begin{align}
  M_1 &= 11 \frac{g_Y^2}{16 \pi^2} m_{3/2} \left[ 1 - (\alpha -1)^2 -  \frac{ t(\gamma + t + \gamma t)}{1 + \gamma t}  \right], \notag \\
  M_2 &=  \frac{g_2^2}{16 \pi^2} m_{3/2} \left[1 - 5 (\alpha-1)^2 -5 \frac{ t (\gamma + t + \gamma t^2)}{1 + \gamma t} \right], \notag \\
  M_3 &= - 3 \frac{g_3^2}{16 \pi^2} m_{3/2} \left[ 1 + (\alpha - 1)^2 + \frac{ t (\gamma + t + \gamma t^2) }{1 + \gamma t} \right]. \label{m1m2m3}
  \end{align}
  For example, if we choose $\gamma=1$ (as in eq. (\ref{parameterchoice})) the above equations give
  \begin{align}   M_1 & \approx 0.05 \  g_Y^2 \  m_{3/2}, \notag \\   M_2 & \approx 0.048 \ g_2^2 \ m_{3/2}, \notag \\   M_3 & \approx 0.052 \ g_3^3 \  m_{3/2}.   \end{align}
These relations are compatible accidentally with gauge coupling unification. Indeed, if we now assume that the gauge couplings unify at some unification scale $\frac{5}{3} g_Y^2\equiv g_1^2 = g_2^2 = g_3^2 = 0.51$, we get the gaugino masses at this scale 
\begin{align} M_1 & \approx 0.015\ m_{3/2}, \notag \\  M_2 & \approx 0.025\ m_{3/2},\notag \\ M_3 & \approx 0.026\ m_{3/2}. \end{align}
The gaugino masses for other values of $\gamma$ are listed in table \ref{tanbeta} below.

Note that in a similar way, the trilinear terms $A_0$ also receive corrections proportional to   
\begin{align}   \delta A_{ijk} = - \frac{1}{2} \left( \gamma_i + \gamma_j + \gamma_k \right) m_{3/2},   \end{align}  where the $\gamma$'s are the anomalous dimensions of the corresponding cubic term in the superpotential. These contributions however are small compared to the tree-level value in eq. (\ref{softterms}).
  
Although the gaugino masses are generated at one-loop, our model is very different from a mAMSB (minimal Anomaly Mediated Supersymmetry Breaking) \cite{gauginomass} scenario: In mAMSB, the second and third term in eq. (\ref{gaugino mass}) are missing due to the absence of hidden sector fields with a Planck scale VEV.  In our model however, the second term in eq. (\ref{gaugino mass}) is present because of the non-vanishing F-terms of the $s$ and $z$ fields,  and has the effect that it raises the gaugino masses slightly to the order $M_{1/2} \approx 2 \times 10^{-2}  \ m_{3/2}$ compared to $M_{1/2} \approx 10^{-2} - 10^{-3} \ \ m_{3/2}$ for a mAMSB where only the first term in eq. (\ref{gaugino mass}) is non-vanishing. Another important difference is that we have $M_{1} < M_{2}$ which results in a mostly Bino-like LSP (Lightest Supersymmetric Particle), compared with a mostly Wino-like LSP in mAMSB. Note also that we do not have any danger of tachyonic scalar soft masses because of the presence of a tree-level soft mass $m_0$ in eq. (\ref{softterms}).  We also have tree-level trilinear couplings $A_0$, which are not present in the mAMSB.

Our model is also different from the minimal supergravity mediated scenario (mSUGRA) \cite{mSUGRA}. Indeed, in mSUGRA gaugino masses are imposed to be equal at tree-level at the GUT unification scale $M_3 \ : \ M_2 \ : \ M_1 = g_3^2 \ : \ g_2^2 \ : \ g_1^2$ of the order $m_0$ (plus or minus an order of magnitude), while our model has vanishing tree-level gaugino masses. They are generated at one-loop and do not satisfy the above relation. Since the gaugino masses are generated at one-loop they are much smaller than the other soft terms.

We conclude that although the soft terms $m_0, A_0$ and $B_0 = A_0 - m_{3/2}$ are similar to an mSUGRA scenario, the anomaly mediated gaugino masses  (which have on top of the usual AMSB contribution proportional to the beta function another contribution from the Planck scale VEVs of $s$ and $z$) are not universal and are much smaller.  Therefore, the particle spectrum will resemble much more the spectrum of a mAMSB scenario, with the important difference that the lightest neutralino is Bino-like instead of Wino-like (See section \ref{sec:pheno}). 
  
  \section{K\"ahler transformation and gaugino masses} \label{sec:modelR}

  In this section we show that the gaugino masses of the model (\ref{modelMSSM}) and of the model obtained after a K\"ahler transformation (\ref{KahlertransformedMSSM}) match.   While in the first the gaugino masses are generated at one-loop by eq. (\ref{gaugino mass}), the second receives an extra contribution due to a field-dependent part in the gauge kinetic functions which is needed to cancel the mixed $U(1)_R \times G$ anomalies by a Green-Schwarz counter term. The anomalous contributions to the gauge transformations are proportional to $\mathcal C_A$, given by
  \begin{align}\label{CA}     \mathcal C_A \delta^{ab} &= \Tr \left[ R_\psi (\tau^a \tau^b )_A \right] + T_{G_A} \delta^{ab} R_\lambda\, ,  \end{align}
  where $A=Y,2,3$ labels the Standard Model gauge groups. 
  The R-charge of the matter fermions is $R_\psi = bc/2$, while the gauginos carry a charge $R_\lambda = -bc/2$, such that eq.~(\ref{CA}) can be rewritten as
  \begin{align}   \mathcal C_A = \frac{bc}{2} \left(T_{R_A} - T_{G_A} \right).   \end{align}
   Anomaly cancellation (as in eq. (\ref{anomalieformules})) then requires that
  \begin{align}   \beta_A = \frac{  \mathcal C_A} {4 \pi^2 c } \label{anomalymatching}   \end{align}
  The effect of these (quantum) corrections to the gauge kinetic functions compared to the classically equivalent theory in eq. (\ref{modelMSSM}) is that non-zero gaugino masses $m_R$ for the R-gaugino and for the Standard Model gauginos $m_A$ are now generated because of a field-dependent gauge kinetic function, on top of  the ``anomaly mediation" contribution (\ref{gaugino mass}).   The corresponding contribution to the gaugino masses can be calculated using eq. (\ref{mgaugino}) together with the anomaly matching conditions eqs. (\ref{anomalymatching}).   
  \begin{align}   m_{A} &= -\frac{g_A^2}{2} e^{\kappa^2 \mathcal K/2} \beta_A g^{\alpha \bar \beta} \bar \nabla_{\bar \beta} \bar W \notag \\   &= \frac{g_A^2}{16 \pi^2} b (T_G - T_R)  e^{\kappa^2 \mathcal K/2} g^{\alpha \bar \beta} \bar \nabla_{\bar \beta} \bar W\, , \label{gauginoRmass1}   \end{align}
  where it is taken into account that the masses of the MSSM gauginos calculated by (\ref{mgaugino}) need a rescaling proportional to $g_A^2$ due to their non-canonical kinetic terms:
  \begin{align}   \mathcal L/e &= -\frac{1}{2} \text{Re}(f)_A \bar \lambda^A \cancel D \lambda^A \notag \\   &= -\frac{1}{2} \left(\frac{1}{g_A^2} + \beta_A \frac{\alpha}{b} \right)  \bar \lambda^A \cancel D \lambda^A, \label{gauginoRmass}   \end{align}
  where $\beta_A \frac{\alpha}{b} < < g_A^{-2}$ if the gauge coupling is in the perturbative region. 
 
  On the other hand, since the K\"ahler potential differs by a linear term $b(s + \bar s)$, the contribution of the second term in  eq. (\ref{gaugino mass}) differs by a factor
  \begin{align}   \delta m_{A} =  \frac{g_A^2}{16 \pi^2} (T_G - T_R) b e^{\kappa^2 \mathcal K/2} g^{\alpha \bar \beta} \bar \nabla_{\bar \beta} \bar W,   \end{align}
  which exactly coincides with eq. (\ref{gauginoRmass1}). 

  We conclude that even though the models (\ref{modelMSSM}) and (\ref{KahlertransformedMSSM}) differ by a (classical) K\"ahler transformation, they generate the same gaugino masses at one-loop given in eq. (\ref{m1m2m3}).   While the one-loop gaugino masses for the model (\ref{modelMSSM}) are generated entirely by eq. (\ref{gaugino mass}), the gaugino masses for the model (\ref{KahlertransformedMSSM}) after a K\"ahler transformation    have a contribution from eq. (\ref{gaugino mass}) as well as from a field dependent gauge kinetic term whose presence is necessary to cancel the mixed $U(1)_R \times G$ anomalies due to the fact that the extra $U(1)$ has become an R-symmetry giving an R-charge to all fermions in the theory.    
  
  \section{Phenomenology} \label{sec:pheno}

  The results for the soft terms calculated in section \ref{sec:extrahidden}, evaluated for different values of the parameter $\gamma$ are summarised in table \ref{tanbeta}.
  For every $\gamma$, the corresponding $t$ and $\alpha$ are calculated by imposing a vanishing cosmological constant by eqs. (\ref{ca}) and (\ref{ca2}).
  The scalar soft masses and trilinear terms are then evaluated by eqs. (\ref{softterms}) and the gaugino masses by eqs. (\ref{m1m2m3}). 
  Note that the relation (\ref{AB-relation}), namely $ A_0 = B_0 - m_{3/2} $, is valid for all $\gamma$. We therefore do not list the parameter $B_0$.

\begin{table}[h!] \centering \begin{tabular}{| l | ll | lllll | rr |} \hline
$\gamma$ & t     & $\alpha$ & $m_0$ & $A_0$ & $M_1$ & $M_2$ & $M_3$ & $\tan \beta (\mu> 0)$ & $\tan \beta (\mu < 0)$ \\  \hline 
0.6      & 0.446 & -0.175   & 0.475         & 1.791         & 0.017         & 0.026         & 0.027         &                                   &                               \\
1        & 0.409 & -0.134   & 0.719         & 1.719         & 0.015         & 0.025         & 0.026         &                                   &                                \\
1.1      & 0.386 & -0.120   & 0.772         & 1.701         & 0.015         & 0.024         & 0.026         & 46                                & 29                             \\
1.4      & 0.390 & -0.068   & 0.905         & 1.646         & 0.014         & 0.023         & 0.026         & 40                                & 23                             \\
1.7      & 0.414 & -0.002   & 0.998         & 1.588         & 0.013         & 0.022         & 0.025         & 36                                & 19                              \\ \hline 
\end{tabular} \caption{The soft terms (in terms of $m_{3/2}$) for various values of $\gamma$. If a solution to the RGE exists, the value of $\tan \beta$ is shown in the last columns for $\mu >0$ and $\mu <0$ respectively.} \label{tanbeta} \end{table} 

In most phenomenological studies, $B_0$ is substituted for $\tan \beta$, the ratio between the two Higgs VEVs, as an input parameter for the renormalization-group equations (RGE) that determine the low energy spectrum of the theory. Since $B_0$ is not a free parameter in our theory, but is fixed by eq. (\ref{AB-relation}), this corresponds to a definite value of $\tan \beta$. For more details see \cite{Ellis} (and references therein). The corresponding $\tan \beta$ for a few particular choices for $\gamma$ are listed in the last two columns of table \ref{tanbeta} for $\mu>0$ and $\mu<0$ respectively. No solutions were found for $\gamma \lesssim 1.1$, for both signs of $\mu$. 

\begin{figure}[!h]  \centering  \begin{minipage}[b]{0.45\textwidth}  \includegraphics[width=\textwidth]{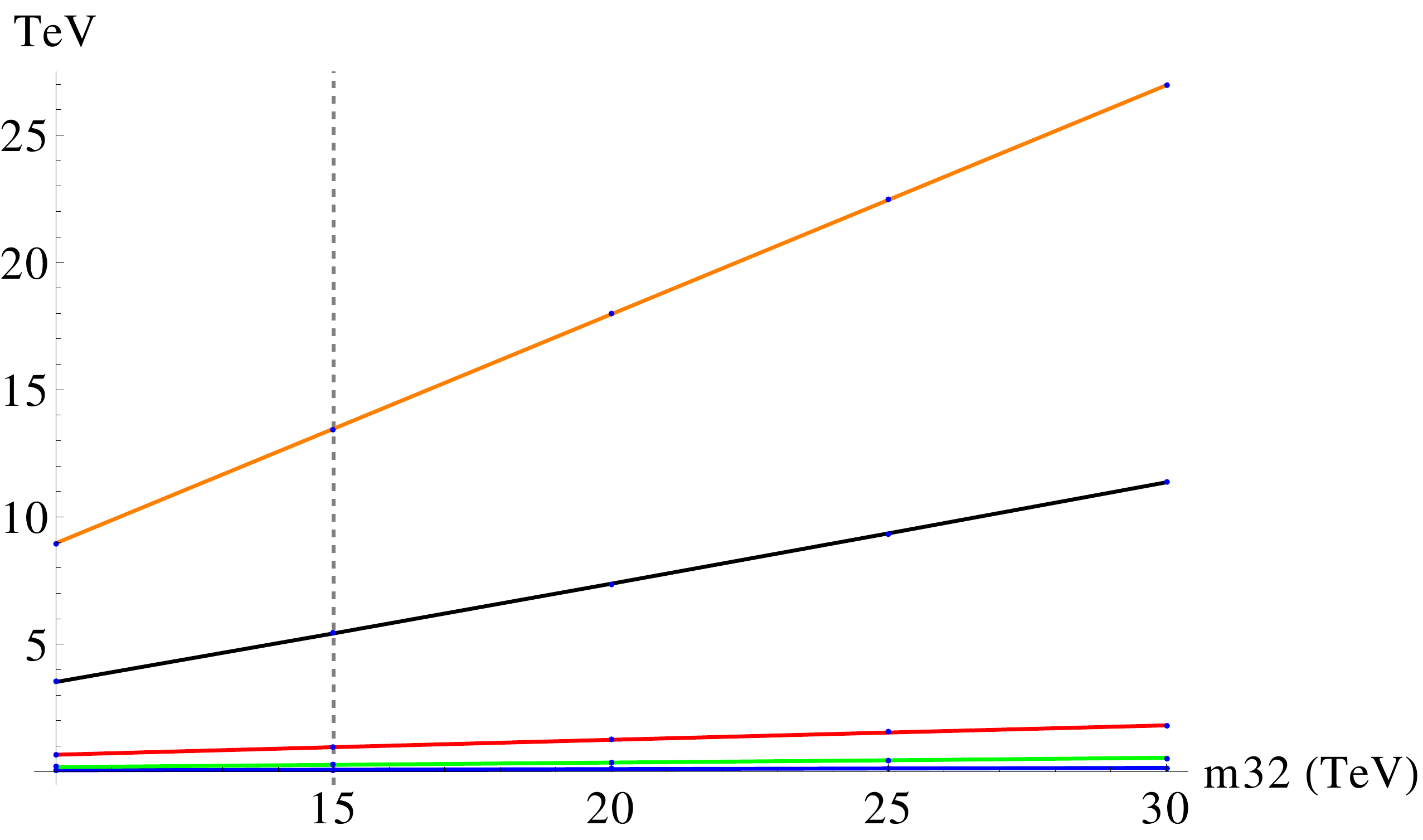}  \caption*{$\mu>0$} \end{minipage} \hfill
\begin{minipage}[b]{0.45\textwidth}  \includegraphics[width=\textwidth]{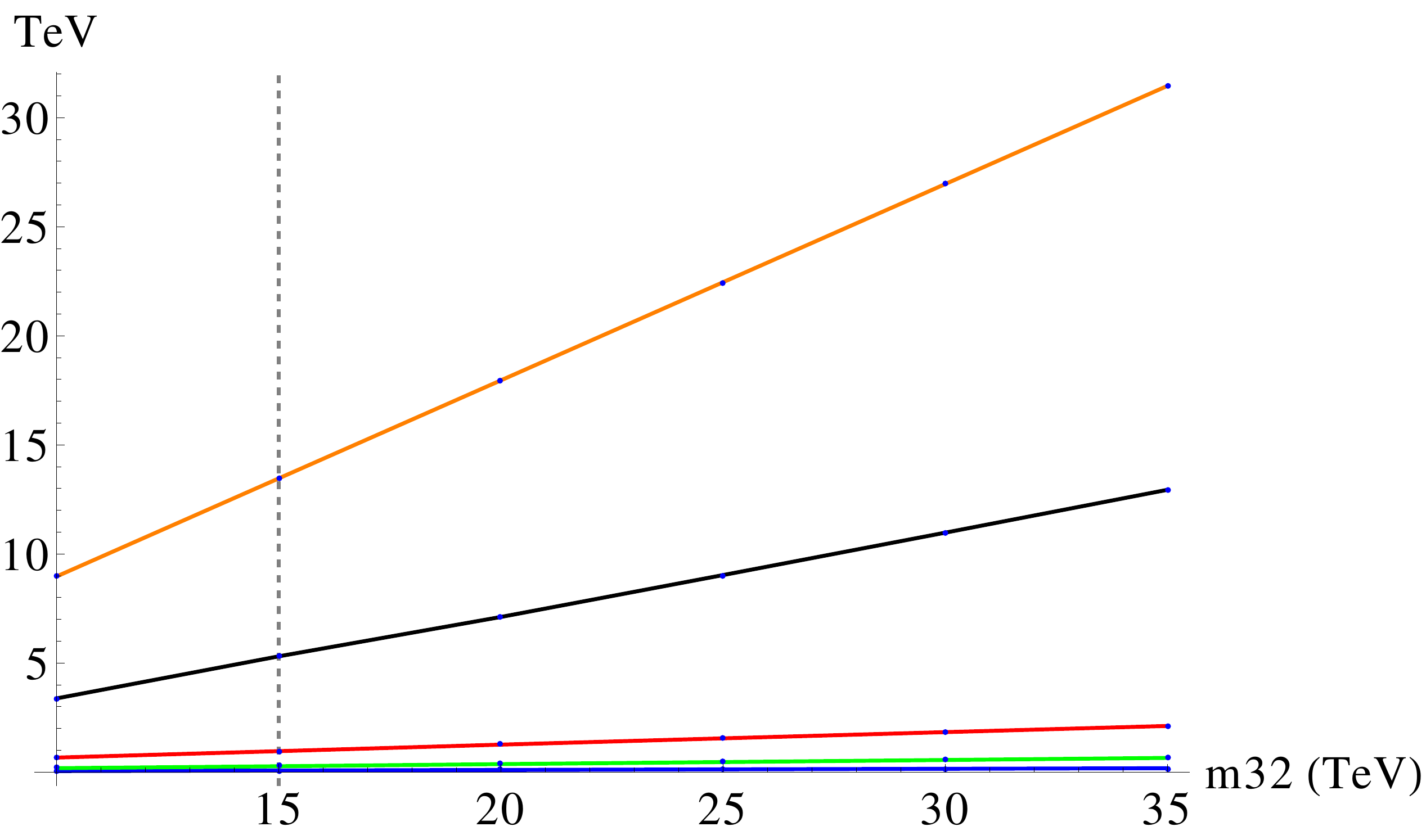} \caption*{$\mu<0$} \end{minipage}
\caption{The masses (in TeV) of the sbottom squark (yellow), the stop squark (black), the gluino (red), the lightest chargino (green) and the lightest neutralino (blue) as a function of the gravitino mass for $\gamma=1.4$ and for $\mu >0$ (left) and $\mu<0$ (right).
The mass of the lightest neutralino varies slightly between 42 GeV (46 GeV) for $m_{3/2}=10$ TeV and 138 GeV (149 GeV) for  $m_{3/2}=30$ TeV for $\mu>0$ ($\mu<0$). The vertical dashed line at $m_{3/2 \approx 15}$ TeV indicates the exclusion limit (lower bound) on the gluino mass. }\label{plotm32} \end{figure}

Some characteristic masses \cite{softsusy} for $\gamma=1.4$ as a function of the gravitino mass are shown in figure \ref{plotm32}. 
A lower experimental bound of 1 TeV for the gluino mass (vertical dashed line) forces $m_{3/2} \gtrsim 15$ TeV. On the other hand, for $\mu > 0$ ($\mu <0$) no viable solution for the RGE was found 
when $m_{3/2}\gtrsim 30$ TeV ($m_{3/2}\gtrsim 35$ TeV). 
We conclude that (for $\gamma = 1.4$) 
\begin{align} & 15 \text{ TeV} \lesssim m_{3/2} \lesssim 30 \text{ TeV} && \text{for }\mu > 0, \notag \\ & 15 \text{ TeV} \lesssim m_{3/2} \lesssim 35 \text{ TeV} && \text{for }\mu < 0. \end{align}
As we will see below, these upper bounds can differ for different choices of $\gamma$. 

\begin{figure}[h!]  \centering \begin{minipage}[b]{0.45\textwidth}  \includegraphics[width=\textwidth]{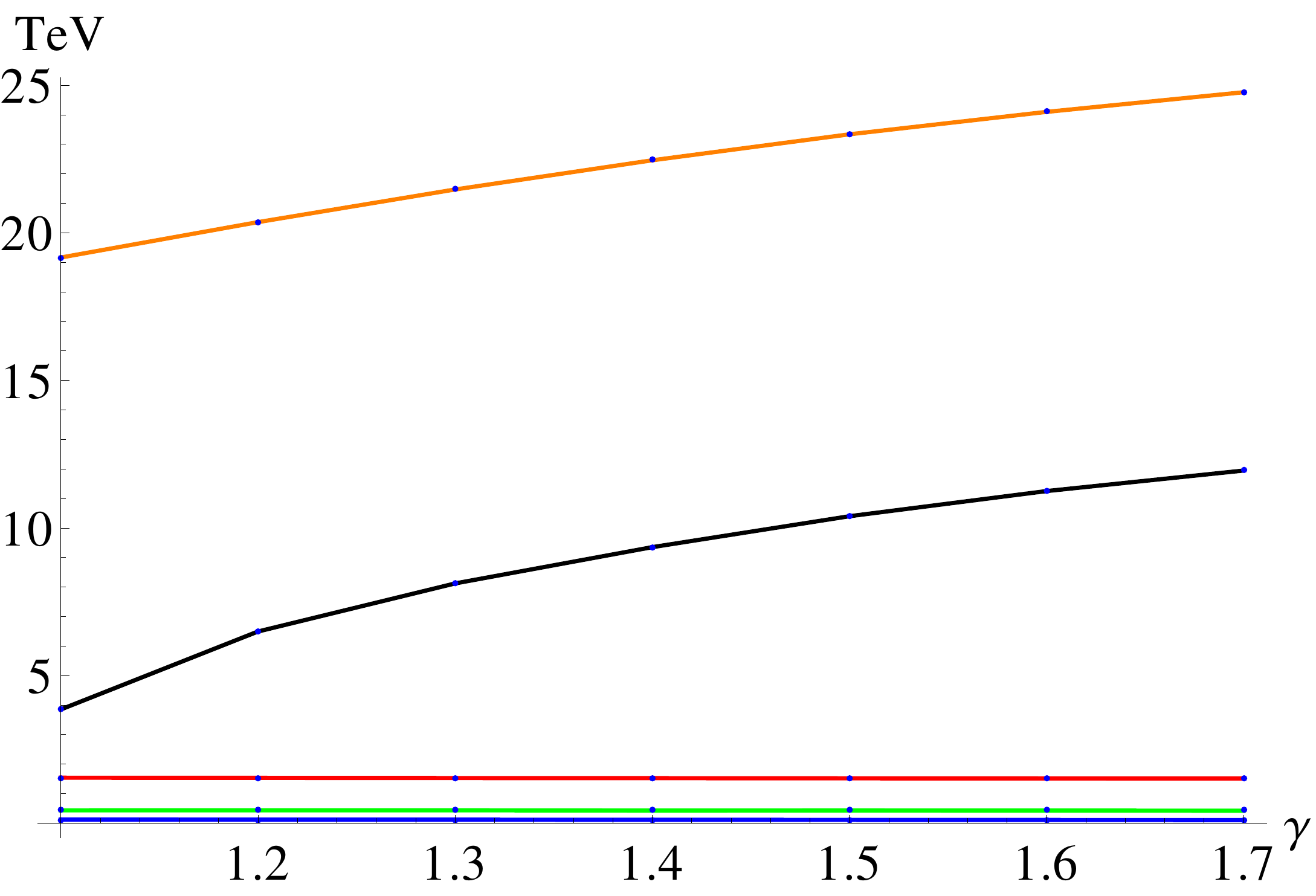} \caption*{$\mu>0$} \end{minipage} \hfill
\begin{minipage}[b]{0.45\textwidth}  \includegraphics[width=\textwidth]{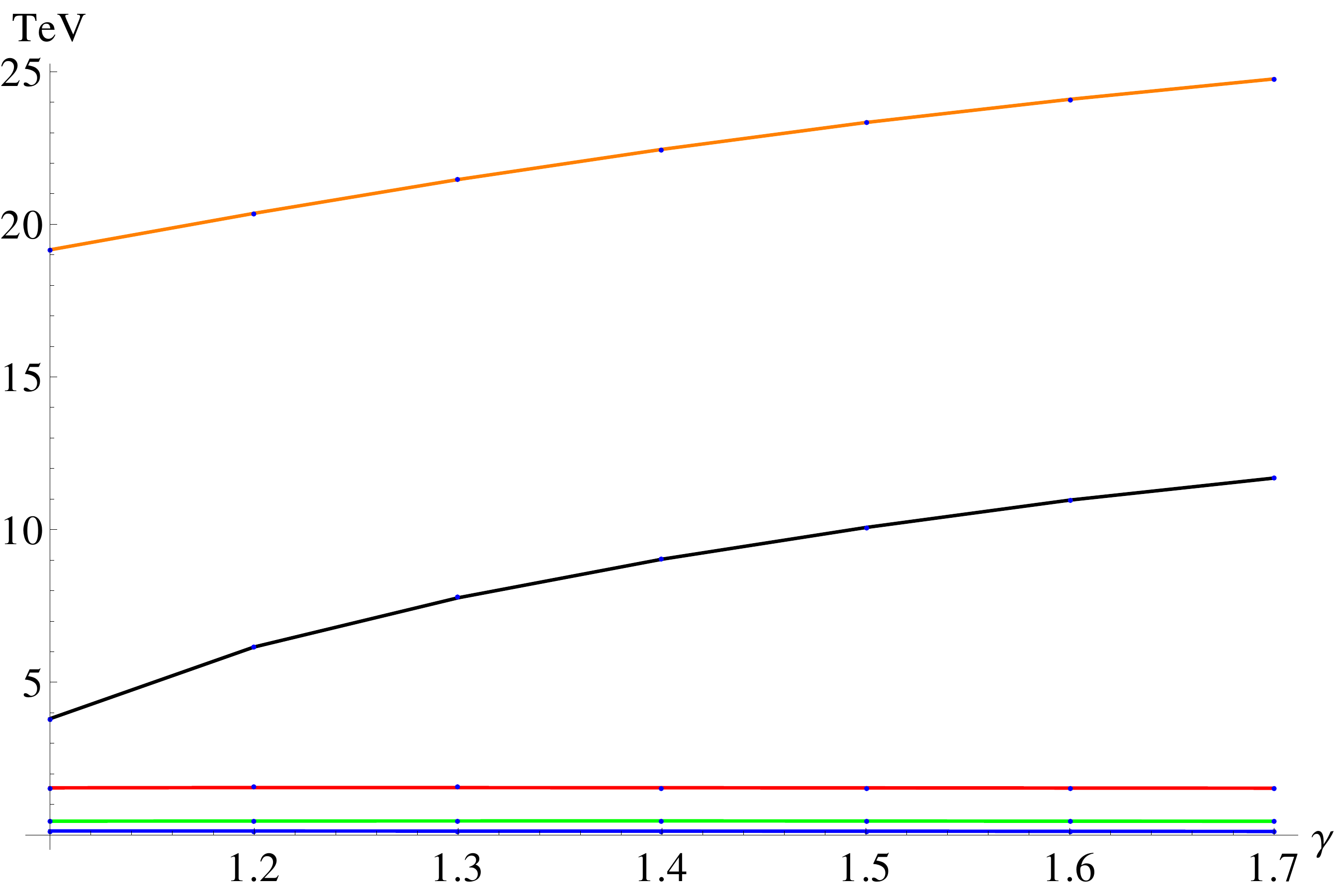} \caption*{$\mu<0$} \end{minipage}
\caption{The masses (in TeV) of the sbottom squark (yellow), the stop squark (black), the gluino (red), the lightest chargino (green) and the lightest neutralino (blue) as a function of $\gamma$ for $m_{3/2}=25$ TeV and for $\mu >0$ (left) and $\mu<0$ (right). No solutions for the RGE were found for $\gamma <1.1$. Notice that for $\gamma \rightarrow 1.1$ the stop mass becomes relatively light.} \label{plotgamma} \end{figure}

In figure \ref{plotgamma}, the same spectrum is plotted  as a function of $\gamma$ for $m_{3/2}=25$ TeV. As one can see, the stop mass varies heavily with $\gamma$, and can become relatively light when $\gamma \approx 1.1$.
For all values of $\gamma$ the LSP is given by the lightest neutralino and since $M_1 < M_2$ (see table \ref{tanbeta}) the lightest neutralino is mostly Bino-like, in contrast with a typical mAMSB scenario, where the lightest neutralino is mostly Wino-like \cite{winolike}.

\begin{figure}[h!]  \centering \begin{minipage}[b]{0.45\textwidth}  \includegraphics[width=\textwidth]{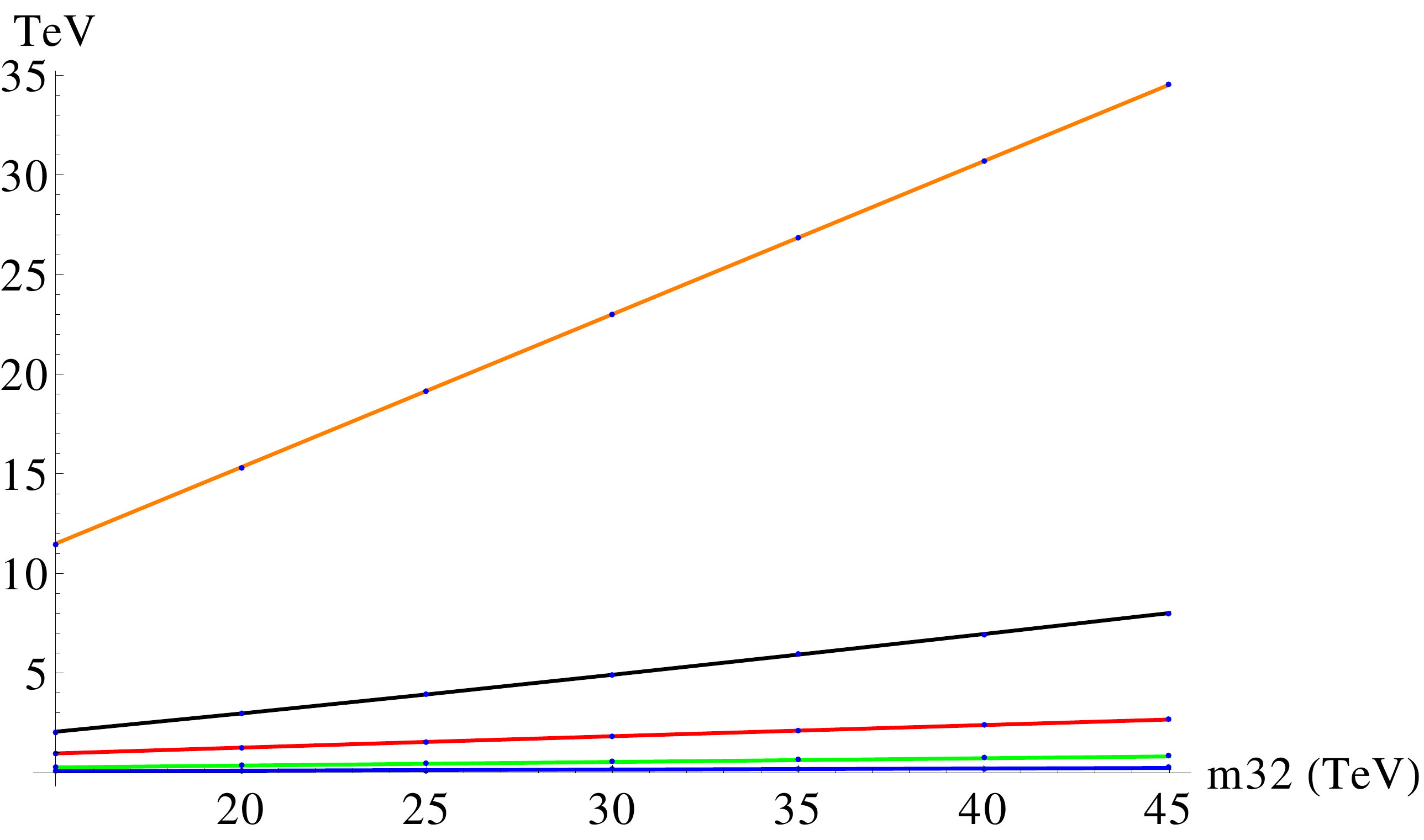} \caption*{$\gamma=1.1$ and $\mu >0$} \end{minipage} \hfill
\begin{minipage}[b]{0.45\textwidth}  \includegraphics[width=\textwidth]{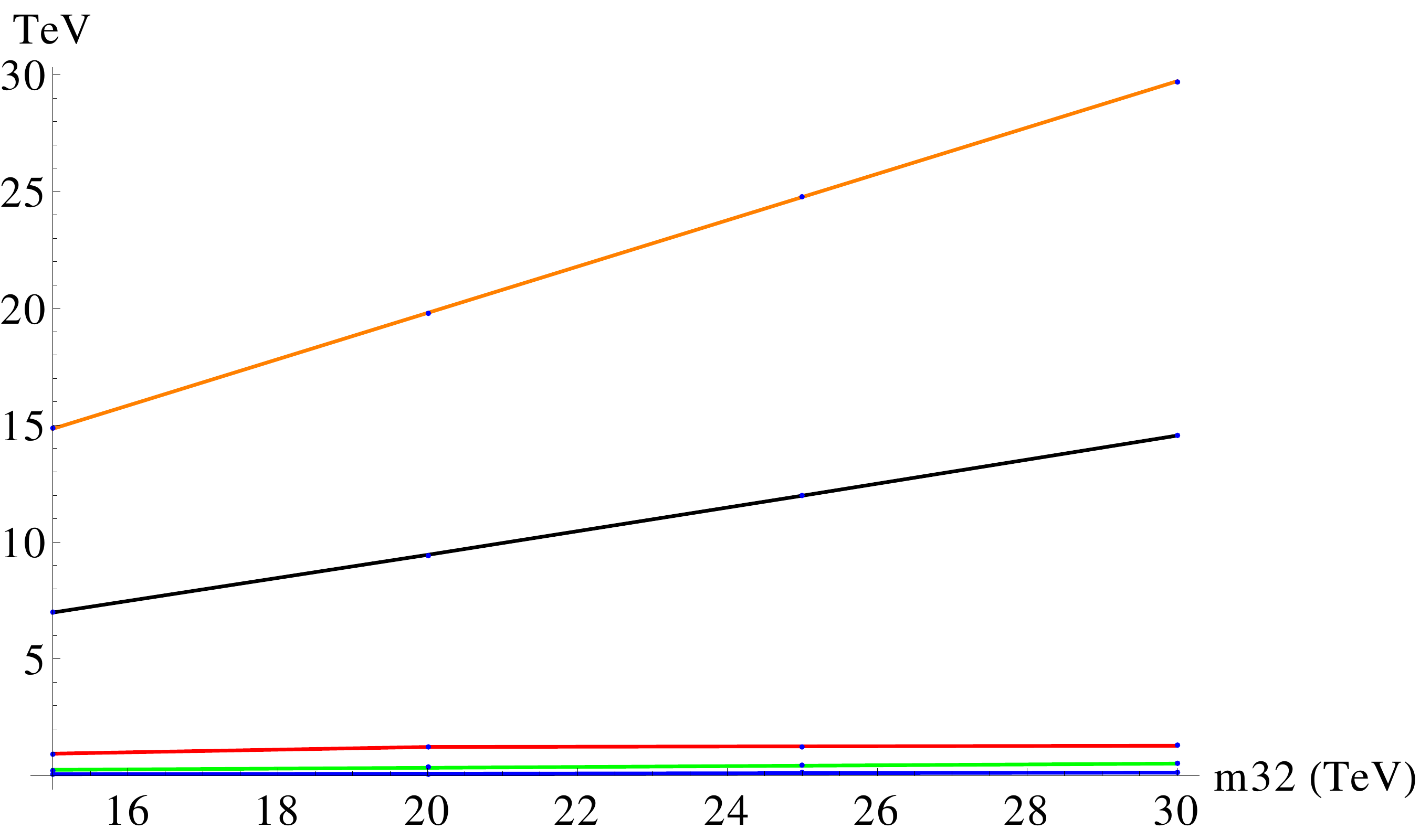} \caption*{$\gamma=1.7$ and $\mu>0$} \end{minipage}
\caption{The masses (in TeV) of the sbottom squark (yellow), the stop squark (black), the gluino (red), the lightest chargino (green) and the lightest neutralino (blue) as a function of $m_{3/2}$ for $\gamma=1.1$ (left) and for $\gamma=1.7$ (right), for $\mu>0$. 
For $\gamma=1.1$ (left) no solutions to the RGE were found when $m_{3/2} \gtrsim 45$ TeV, while for $\gamma=1.7$ (right) no solutions were found when $m_{3/2} \gtrsim 30$ TeV. The lower bound corresponds in both cases to a gluino mass of 1 TeV. } \label{fig_11} \end{figure}

To get a lower bound on the stop mass, the sparticle spectrum is plotted in figure \ref{fig_11} (left) as a function of the gravitino mass for $\gamma=1.1$ and $\mu >0$ (for $\mu <0$ the bound is higher).
As above, the experimental limit on the gluino mass forces $m_{3/2} \gtrsim 15$ TeV. In this limit the stop mass can be as low as 2 TeV.
To obtain an upper bound on the stop mass on the other hand, the sparticle spectrum is plotted in figure \ref{fig_11} (right) for $\gamma =1.7$ and $\mu>0$. 
Above a gravitino mass of (aproximately) 30 TeV, no solutions to the RGE were found. In this limit the stop mass is about 15 TeV. 

To conclude, the lower end mass spectrum consists of (very) light charginos (with a lightest chargino between 250 and 800 GeV) and neutralinos, with a mostly Bino-like neutralino as LSP ($80-230$ GeV), which would distinguish this model from the mAMSB where the LSP is mostly Wino-like.
These upper limits on the LSP and the lightest chargino imply that this model could in principle be excluded in the next LHC run.
In order for the gluino to escape experimental bounds, the lower limit on the gravitino mass is about 15 TeV. The gluino mass is then between 1-3 TeV. This however forces the squark masses to be very high ($10-35$ TeV), with the exception of the stop mass which can be relatively light ($2-15$ TeV).

 \section{Non-canonical K\"ahler potential for the visible sector} \label{sec:noncan}

Since the model (\ref{model0}) has tachyonic soft scalar masses for the MSSM fields, in section \ref{sec:extrahidden} we proposed a solution by adding an extra field to the hidden sector. However, we will show in this section that the problem can also be circumvented by allowing non-canonical kinetic terms for the MSSM fields.

 We consider the following model
 \begin{align} \mathcal K &= - \kappa^{-2} \log (s + \bar s) + \kappa^{-2} b (s + \bar s) + (s + \bar s)^{-\nu} \sum \varphi \bar \varphi , \notag \\ W &= \kappa^{-3} a + W_{MSSM} , \notag \\  f(s) &= 1, \ \ \ \ \ \ \ f_A(s)=1/g_A^2 \label{model_noncan} \end{align}
 where a sum over all visible sector fields $\varphi$ is understood in the K\"ahler potential. Here, $\nu$ is considered to be an additional parameter in the theory, where $\nu=1$ corresponds with the leading term in the Taylor expansion of $-\log(s + \bar s - \varphi \bar \varphi)$. The gauge kinetic functions for the Standard Model gauge groups $f_A(s)$ are taken to be constants.

The scalar potential is given by
  \begin{align} V &= V_F + V_D, \notag \\ 
V_F &= \kappa^{-4}  \frac{e^{b(s + \bar s) + \sum \kappa^2 \left( s + \bar s\right)^{-\nu} \varphi \bar \varphi}}{s + \bar s} 
\left(-3W\bar W + g^{s \bar s} \left|\nabla_s W \right|^2+ \sum_\varphi (s + \bar s)^{\nu} \left| \nabla_\varphi W \right|^2 \right), \notag \\  
V_D &= \frac{c^2}{2} \left(b - \frac{1}{s + \bar s} - \nu (s  + \bar s)^{-\nu -1} \sum \varphi \bar \varphi \right)^2, \label{scalarpotnoncan}\end{align}
where
\begin{align}   \nabla_\alpha W  = \partial_\alpha W + \kappa^2 (\partial_\alpha \mathcal K) W. \end{align}

Since the visible sector fields appear only in the combination $\varphi \bar \varphi$, their VEVs vanish provided that the scalar soft masses squared are positive.
Moreover, for vanishing visible sector VEVs, the scalar potential reduces to eq. (\ref{scalarpot0}) and the non-canonical K\"ahler potential for the visible sector fields does not change the discussion on the minimization of the potential in section \ref{sec:model}. 
Therefore, the non-canonical K\"ahler potential does not change the fact that the F-term contribution to the soft scalar masses squared is negative. One has as in eq. (\ref{tach})
\begin{align}
 \left. \frac{\partial^2 V_F}{\partial \varphi \partial \bar \varphi} \right|_{\langle \varphi \rangle = 0} = \kappa^{-2}a^2 e^{\alpha} \left(\frac{b}{\alpha} \right)^{\nu + 1} (\langle \sigma_s \rangle  +1) <0,
\end{align}
However, the visible fields will enter in the D-term scalar potential through the derivative of the K\"ahler potential with respect to $s$. 
Even though this has no effect on the ground state of the potential, the $\varphi$-dependence of the D-term scalar potential does result in an extra contribution to the scalar masses squared
\begin{align}
 \left. \frac{\partial^2 V_D}{\partial \varphi \partial \bar \varphi} \right|_{\langle \varphi \rangle = 0} = \nu \kappa^{-2} c^2 \left( \frac{b}{\alpha} \right)^{\nu + 2} \left(1-\alpha \right).
\end{align}
The total soft mass squared is then the sum of these two contributions
\begin{align}
m_0^2 &=  \kappa^2 a^2  \left(\frac{b}{\alpha} \right) \left( e^\alpha (\sigma_s +1) + \nu \frac{A(\alpha)}{\alpha} (1-\alpha)  \right),
\end{align}
where eq. (\ref{bsalpha}) has been used to relate the constants $a$ and $c$, and corrections due to a small cosmological constant have been neglected. A field redefinition due to a non-canonical kinetic term $g_{\varphi \bar \varphi} = (s + \bar s)^{-\nu}$ is taken into account. The soft mass squared is now positive if
\begin{align}  \nu > -\frac{e^\alpha (\sigma_s +1) \alpha }{ A(\alpha) (1-\alpha)} \approx 2.6. \end{align}

The gravitino mass is given by
\begin{align}
 m_{3/2} = \kappa^{-1}a \sqrt{b/\alpha} e^{\alpha/2}.
\end{align}
In the hidden sector, the imaginary part of $s$ is eaten by the gauge boson corresponding to the shift symmetry, which becomes massive (similar to eq. (\ref{gaugebosonmass}))
 \begin{align} m_{A_\mu} = \frac{\kappa^{-1} b c }{\alpha} \approx 1.39 \ m_{3/2}.  \end{align}
The mass of the real part of $s$ squared is given by $2 ( \alpha/b)^2 \partial_s \partial_s V$ evaluated at the ground state, where the factor $2(\alpha/b)^2$ comes from the non-canonical kinetic term,
\begin{align}
 m_s^2 &=2 \left(\alpha ^4-2 \alpha ^2+4 \alpha +\frac{e^{-\alpha } (3-2 \alpha ) A}{\alpha }-4\right)  m_{3/2}^2 \notag \\ &\approx 3.48 \ m_{3/2}^2.
\end{align}
Finally, the Goldstino is given by a linear combination of the fermionic superpartner of $s$ and the gaugino, which is eaten by the gravitino by the BEH mechanism. The mass of the remaining fermion is given by (see Appendix \ref{Appendix:Fermions})
\begin{align}
 m_f^2 &\approx 3.81 \  m_{3/2}^2.
\end{align}

  Note that in the scalar potential eq. (\ref{scalarpotnoncan}) the MSSM F-terms $\sum_\varphi \left| \nabla_\varphi W \right|^2 $ come with a prefactor
$e^{\kappa^2 \mathcal K} g^{\varphi \bar \varphi} $  (where the hidden fields are replaced by their VEVs). To bring this into a more recognizable (globally supersymmetric) form where $\mathcal L \sim - g_{\varphi \bar \varphi} \partial_\mu \varphi \partial^\mu \bar \varphi - g^{\varphi \bar \varphi} W_\varphi \bar W_{\bar \varphi}$, 
one should rescale the MSSM superpotential (defined in eq. (\ref{MSSMsuperpot}))
  \begin{align}
  \hat W_{MSSM} = \exp (\alpha) \left(b/\alpha \right) \ W_{MSSM}.  \label{MSSMhatnoncan}
  \end{align}
However, another rescaling is needed to take into account the non-canonical K\"ahler potential for the visible sector\footnote{
After the rescaling (\ref{MSSMhatnoncan}), the Lagrangian contains (very schematically) the following terms
\begin{align}
 \mathcal L = & -(s + \bar s)^{-\nu} \partial_\mu \bar \varphi \partial^\mu \varphi  -(s + \bar s)^{-\nu} \partial_\mu \bar h \partial^\mu  h + \hat \mu^2 \bar h h + \hat y \hat \mu \bar h \varphi \varphi + \hat y^2 \bar \varphi \bar \varphi \varphi \varphi + \dots \notag \\
 & + \frac{1}{6} A_0 \hat y \varphi \varphi \varphi + \frac{1}{2} B_0 \hat \mu h h.
\end{align}
where $h$ stands for the Higgsinos and $\varphi$ labels the other scalar superpartners and all indices are surpressed for clarity. $y$ stands for the Yukawa couplings and $\mu$ is the usual $\mu$-parameter. The first line contains the kinetic terms and the F-terms coming from $\hat W_{MSSM}$. The last line contains the trilinear supersymmetry breaking terms (A-terms) and the B-term.
In order to obtain canonical kinetic terms, one needs a rescaling $\varphi \rightarrow \varphi'=(s+ \bar s)^{-\nu/2} \varphi$ (and similarly for $h$). However, to bring the MSSM superpotential back into its usual form one also needs to redefine $\hat \mu \rightarrow \hat \mu'=(s+\bar s)^{\nu/2} \hat \mu$ and $\hat y \rightarrow \hat y'=(s+\bar s)^{\nu} \hat y $.
One then obtains
\begin{align}
 \mathcal L = & - \partial_\mu \bar \varphi ' \partial^\mu \varphi '  - \partial_\mu \bar h' \partial^\mu h ' + \hat \mu '^2 \bar h' h' + \hat y' \hat \mu ' \bar h' \varphi ' \varphi ' + \hat y'^2 \bar \varphi ' \bar \varphi ' \varphi ' \varphi ' + \dots \notag \\
 & + \frac{1}{6} (s + \bar s)^{\nu/2} A_0 \hat y ' \varphi ' \varphi ' \varphi ' + \frac{1}{2} (s + \bar s)^{\nu/2} B_0 \hat \mu ' h' h'.
\end{align}
}.  The trilinear couplings are given by 
  \begin{align}     &A_0= m_{3/2}(s + \bar s)^{\nu/2} \left(\sigma_s + 3 \right),   \end{align}
 and
  \begin{align}  &B_0 = m_{3/2} (s + \bar s)^{\nu/2} \left(\sigma_s + 2 \right),  \end{align}
  
The main phenomenological properties of this model are not expected to be different from the one we analyzed in section \ref{sec:pheno} with the parameter $\nu$ replacing $\gamma$. Gaugino masses are still generated at one-loop level while mSUGRA applies to the soft scalar sector. We therefore do not repeat the phenomenological analysis for this model. 

\section{Conclusions}

In this work, we studied a simple supergravity model that allows for an infinitesimally small value of the cosmological constant, while leaving the supersymmetry breaking scale as an independent parameter.

The minimal model contains a single chiral multiplet $S$ (a dilaton) which has a gauged shift symmetry, and a vector multiplet. Supersymmetry breaking is then realised by an expectation value of both an F and D-term. 

A K\"ahler potential of the form $K=-p\log(s + \bar s)$ is assumed, while the most general superpotential is a single exponential. By performing a K\"ahler transformation the exponential superpotential can be absorbed in a linear term in the K\"ahler potential and one is left with a constant superpotential. Gauge invariance then dictates a constant gauge kinetic term, since otherwise a linear contribution would break the (local) shift symmetry.

We showed that when this model is coupled to the MSSM, it leads to tachyonic scalar soft masses. This can be cured by adding an extra Polonyi-like field, or by allowing for non-canonical kinetic terms of the Standard Model fields, while maintaining the desirable features of the model.

This however introduces an extra parameter $\gamma$ (or $\nu$ in the second case), which turns out to be heavily constrained: $\gamma$ should be in the range $\left[ 1.1, 1.707 \right]$, where the lower bound is to prevent a tachyonic stop squark mass, and the upper bound follows from the tunability of the scalar potential.

Since a K\"ahler transformation can bring the theory from a constant superpotential to a theory with an exponential superpotential where the shift symmetry is a gauged R-symmetry, but with non-trivial gauge kinetic functions, there is an apparent puzzle with the gaugino masses that vanish classically in the first representation but not in the second. Indeed in the latter case all fermions in the theory are charged under $U(1)_R$ leading to anomalies that are cancelled by a Green-Schwarz mechanism due to a gauge kinetic function which is linear in $S$.
However, this also results in non-zero gaugino masses, while in the former case the gaugino masses vanish. 
We have shown that when the 'anomaly mediated' contributions to the gaugino masses are included, the gaugino masses on both sides of the K\"ahler transformation match.

Since the soft SUSY breaking parameter $B_0$ is related to the trilinear coupling by $B_0 = A_0 - m_{3/2}$, the ratio between the two Higgs VEVs $\tan \beta$ is not a free parameter and the model turns out to be very predictive.
The low energy spectrum of the theory consists of  (very) light neutralinos, charginos and gluinos, where the experimental bounds on the (mostly Bino-like) LSP, the lightest chargino and the gluino mass force the gravitino mass to be above 15 TeV.
This in turn implies that the squarks are very heavy, with the exception of the stop squark which can be as light as 2 TeV when the parameter $\gamma$ approaches its lowest limit $\gamma \rightarrow 1.1$.

It follows that the resulting spectrum can be distinguished from other models of supersymmetry breaking and mediation such as mSUGRA and mAMSB.

  \section*{Acknowledgements}
  
  R.K. would like to thank J.P. Derendinger, J. Ellis and F. Zwirner for useful discussions.
  This research was (partly) supported by the NCCR SwissMAP, funded by the Swiss National Science Foundation. 
  
  \appendix

\section{Fermion masses} \label{Appendix:Fermions}

The fermion mass Lagrangian for the chiral fermions $\chi^\alpha$, the gauginos $\lambda^A$ and the gravitino $\psi_\mu$ is given by \cite{VP}
\begin{align} \mathcal L_m = \frac{1}{2} m_{3/2} \bar \psi_\mu P_R \gamma^{\mu \nu} \psi_\nu - \frac{1}{2} m_{\alpha \beta} \bar \chi^\alpha \chi^\beta - m_{\alpha A} \bar \chi^\alpha \lambda^A - \frac{1}{2} m_{AB}\bar \lambda^A P_L \lambda^B + \text{h.c.} \end{align}
where,
\begin{align} m_{\alpha \beta} &= e^{\kappa^2 \mathcal K/2} \left[ \partial_\alpha + (\kappa^2 \partial_\alpha \mathcal K) \right] \nabla_\beta W - e^{\kappa^2 \mathcal K/2} \Gamma^\gamma_{\alpha \beta} \nabla_\gamma W, \notag \\ m_{\alpha A} &= m_{A \alpha} = i \sqrt 2 \left[\partial_\alpha \mathcal P - \frac{1}{4} f_{AB,\alpha} \text{Re}(f)^{-1 \ BC} \mathcal P_C \right] , \notag \\ m_{AB} &= -\frac{1}{2}e^{\kappa^2\mathcal K/2}f_{AB,\alpha} g^{\alpha \bar \beta} \bar \nabla_{\bar \beta} \bar W. \end{align}
Here, $\Gamma^\alpha_{ \beta \gamma} = g^{\alpha \bar \delta} \partial_\beta g_{\gamma \bar \delta}$ is the Christophel connection with as only non-vanishing component $\Gamma^s_{ss} = -\frac{2}{s + \bar s}$. The moment maps $\mathcal P_\alpha$ are defined in eq. (\ref{momentmap}), while $m_{AB} =0$ since the gauge kinetic function is constant.

The Goldstino $P_L \nu$ is given by
\begin{align}  P_L \nu = \chi^\alpha \delta_s \chi_\alpha + P_L \lambda^A \delta_s P_R \lambda_A, \end{align}
where $P_{L(R)}$ is the left-handed (right-handed) projection operator. As before, chiral multiplets are labeled by the index $\alpha$, while the different gauge groups are labeled by the index $A$. The 'fermion shifts' (the scalar parts of the supersymmetry transformation rules) are given by
\begin{align} \delta_s \chi_\alpha &= - \frac{1}{\sqrt 2} e^{\kappa^2 \mathcal K/2} \nabla_\alpha W, \notag \\ \delta_s P_R \lambda_A &= -\frac{i}{2} \mathcal P_A. \end{align}
Due  to  the  super-BEH  effect,  elimination of the  Goldstino  will  give  mass  to  the  gravitino 
\begin{align}  m_{3/2} = \kappa^2 e^{\kappa^2 \mathcal K/2} W. \end{align}
As a result, the mass matrix for the fermions becomes
\begin{align}  m =   \begin{pmatrix}  m_{\alpha \beta} + m_{\alpha \beta}^{(\nu)}  & m_{\alpha B} + m_{\alpha B}^{(\nu)}    \\  m_{A \beta} + m_{A \beta}^{(\nu)}      &     m_{AB} + m_{AB}^{(\nu)}       \end{pmatrix}, \end{align}
where the corrections to the fermion mass terms due to the elimination of the Goldstino are given by
\begin{align} m_{\alpha \beta}^{(\nu)} &= - \frac{4 \kappa^2}{3 m_{3/2}} (\delta_s \chi_\alpha)(\delta_s \chi_\beta), \notag \\  m_{\alpha A}^{(\nu)} &=- \frac{4 \kappa^2}{3 m_{3/2}} (\delta_s \chi_\alpha)(\delta_s P_R \lambda_A ),\notag \\ m_{AB}^{(\nu)} &= - \frac{4 \kappa^2}{3 m_{3/2}} (\delta_s P_R \lambda_A)(\delta_s P_R \lambda_B ). \end{align}

Since the elimination of the Goldstino results in a reduction of the rank of $m$, its determinant vanishes and the physical masses correspond to the non-zero eigenvalues of $m$.

The fermion mass matrix for the model in section \ref{sec:noncan} for the fermionic superpartner of $s$ and the gaugino corresponding to the shift symmetry (\ref{shift}) is then given by.
\begin{align}  m = \kappa^{-1} \left( \begin{array}{cc}   \left( \frac{\alpha}{b}\right)^2 \frac{ a e^{\alpha /2} \left(\alpha ^2+4 \alpha -2\right)}{3 \left(\alpha/b \right)^{5/2}} & - \left( \frac{\alpha}{b}\right) \frac{i \sqrt{2} b^2 c \left(\alpha ^2-2 \alpha -2\right)}{3 \alpha ^2} \\  - \left( \frac{\alpha}{b}\right) \frac{i \sqrt{2} b^2 c \left(\alpha ^2-2 \alpha -2\right)}{3 \alpha ^2} & \frac{c^2 e^{-\frac{\alpha }{2}} (\alpha -1)^2}{3 a \left(\alpha/b \right)^{3/2}} \\ \end{array} \right), \end{align}
Where the factors $\left( \frac{\alpha}{b}\right)$ have been taken into account due to non-canonical kinetic terms for the chiral fermions. The gaugino already has canonical kinetic terms since $f(s)=1$. The hidden sector fermions do not mix with the fermions of the MSSM. Also, the determinant of $m$ is proportional to $(2 + 8 \alpha - 3 \alpha^2 - 2 \alpha^3 + \alpha^4)$, which indeed has a root at $\alpha \approx -0.23315$. The mass squared of the physical fermion is then given by
\begin{align}  m_f^2 &= \left(2\alpha /b \right)^2 \Tr \left[ m^\dagger m \right] = m_{3/2}^2 f_\chi, \end{align}
where 
\begin{align}f_\chi&=  \frac{e^{-2 \alpha } \left(e^{2 \alpha } \alpha ^2 \left(\alpha ^2+4 \alpha -2\right)^2+(\alpha -1)^4 A(\alpha)^2+4 e^{\alpha } \alpha  \left(\alpha ^2-2 \alpha -2\right)^2 A(\alpha)\right)}{9 \alpha ^2} \notag \\ &\approx 3.807, \end{align}
and we have used the relations between the parameters and the numerical values for $\alpha$ and $A(\alpha)$ in eqs. (\ref{bsalpha}).

We now calculate the fermion masses for the model with the extra hidden sector field $z$ in section \ref{sec:extrahidden}. This model contains one extra hidden sector fermion. Its mass matrix is given by $\kappa^{-1}$ times
\begin{align} \notag \hspace{-1.5cm} \left( \begin{array}{ccc}  \left(\frac{\alpha}{b}\right)^2 \frac{ a e^{\frac{1}{2} \left(t^2+\alpha \right)} \left(-2+4 \alpha +\alpha ^2\right) (1+t \gamma )}{3 \left(\frac{\alpha }{b}\right)^{5/2}} & \left(\frac{\alpha}{b}\right) \frac{a e^{\frac{1}{2} \left(t^2+\alpha \right)} (-1+\alpha ) \left(t+\gamma +t^2 \gamma \right)}{3 \left(\frac{\alpha }{b}\right)^{3/2}} & -\left(\frac{\alpha}{b}\right) \frac{i \sqrt{2} b^2 c \left(-2-2 \alpha +\alpha ^2\right)}{3 \alpha ^2} \\  \left(\frac{\alpha}{b}\right) \frac{a e^{\frac{1}{2} \left(t^2+\alpha \right)} (-1+\alpha ) \left(t+\gamma +t^2 \gamma \right)}{3 \left(\frac{\alpha }{b}\right)^{3/2}} & \frac{a e^{\frac{1}{2} \left(t^2+\alpha \right)} \left(2 t \gamma +2 t^3 \gamma -2 \gamma ^2+t^4 \gamma ^2+t^2 \left(1+2 \gamma ^2\right)\right)}{3 \sqrt{\frac{\alpha }{b}} (1+t \gamma )} & -\frac{i \sqrt{2} b c (-1+\alpha ) \left(t+\gamma +t^2 \gamma \right)}{3\alpha ( 1+t  \gamma )} \\  -\left(\frac{\alpha}{b}\right) \frac{i \sqrt{2} b^2 c \left(-2-2 \alpha +\alpha ^2\right)}{3 \alpha ^2} & -\frac{i \sqrt{2} b c (-1+\alpha ) \left(t+\gamma +t^2 \gamma \right)}{3\alpha (1 +t  \gamma )} & \frac{b^2 c^2 \sqrt{\frac{\alpha}{b}} e^{\frac{1}{2} \left(-t^2-\alpha \right)} \left(1-\frac{1}{\alpha }\right)^2}{3 a(1+ t \gamma )} \end{array} \right) \end{align}
It has been checked that the determinant of this matrix vanishes for $\alpha$ and $t$ satisfying eqs. (\ref{ca}) and (\ref{ca2}). The masses of the physical fermions are the two non-zero eigenvalues of this matrix. The result however is quite tedious and we only state the numerical vqlues for $\gamma=1$:
\begin{align}   m_{\chi_1} & \approx  2.57 \ m_{3/2}, \notag \\  m_{\chi_2} & \approx  0.12 \ m_{3/2}. \end{align}

\section{Anomaly cancellation:} \label{Appendix:Anomalies}

In this Appendix we calculate the cubic $U(1)_R^3$ and the mixed $U(1)_R \times G_{\text{SM}}$ anomaly cancellation conditions of the model presented in section \ref{sec:modelR}. In a theory with a gauged R-symmetry, the superpotential transforms under a gauge transformation as $\delta W = -i\xi \theta W$, where $\theta$ is the gauge parameter of the shift symmetry (\ref{shift}), and $\xi = bc$.  Then the charges of all chiral fermions are shifted by $+\xi/2$, so that they become $R_\psi =  \xi/2$. The gauginos and the gravitino have a charge $R_\lambda = -\xi/2$.  The quantum anomalies of such models are studied in full detail in \cite{FreedmanAnomalies,R2}. We summarise their results and apply them to our model.   For the MSSM (fermion) fields, we use the quantum numbers in table \ref{table:charges}.
  \begin{table}[h]   \begin{tabular}{l| lllllll} 	  & Q   & u        & d        & L    & e & $H_u$ & $H_d$ \\ \hline   $U(1)_R$ & $ \xi /2$    & $ \xi /2$       & $ \xi /2$       & $ \xi /2$   & $ \xi /2$ & $\xi /2$     & $\xi /2$     \\   $U(1)_Y$ & 1/6 & -2/3     & 1/3      & -1/2 & 1 & 1/2   & -1/2  \\    $SU(2)$   & 2   & 1        & 1        & 2    & 1 & 2     & 2     \\   $SU(3)$   & 3   & $\bar 3$ & $\bar 3$ & 1    & 1 & 1     & 1        \end{tabular}   \caption{Charge assignments of the various MSSM fermions.} \label{table:charges} \end{table} 
The cubic anomaly is calculated in subsection \ref{Appendix:cubic}. The mixed and gravitational anomalies are calculated in \ref{Appendix:mixed}
  
\subsection{The cubic anomaly} \label{Appendix:cubic}

The one-loop contribution to the gauge transformation $\theta$ from quantum anomalies is given by
\begin{align} \delta \mathcal L_{1-loop} &=  - \frac{\theta}{32 \pi^2} \  \frac{ \mathcal C_R}{3} \ \epsilon^{\mu \nu \rho \sigma} F_{\mu \nu} F_{\rho \sigma} , \notag \\ \mathcal C_R &= \Tr[R_\psi^3] + (n_\lambda + 3) R_\lambda  \label{1-loop} \end{align}
where $n_\chi$ is the number of chiral fermions in the model, $n_\lambda = 8 + 3 + 1 + 1 =13$ is the number of gauginos and the factor '$+3$' comes from the gravitino (3 times the contribution of a gaugino).  The $U(1)_R$ charges $R_\psi$ of the MSSM fields together with their Standard Model gauge group quantum numbers are summarised in table \ref{table:charges}. The trace also includes the hidden sector fields $s$ and $z$ whose R-charge is $R_z = R_s = \xi/2$. We then obtain
\begin{align} \mathcal C_R &= 3 \left[ \left(\frac{\xi}{2} \right)^3 \left(6 + 3+3+2 +1 \right) \right] + \left( \frac{\xi}{2} \right)^3 \left(2+2 \right)  - \left(\frac{\xi}{2} \right)^3 (13 + 3) + 2 \left(\frac{\xi}{2} \right)^3  \notag \\ &=  35 \left( \frac{\xi}{2} \right)^3. \end{align} 
Here, the term in square brackets comes from the MSSM chiral fermions (see table \ref{table:charges}) with a factor 3 for the three different generations of quarks and leptons. The second term in the first line is the contribution from the Higgsinos. The third term is the contribution from the gauginos and the gravitino, while the last term comes from the two hidden sector fields $z$ and $s$.

The one-loop contribution (\ref{1-loop}) is cancelled by a Green-Schwarz mechanism: the Lagrangian contains a term
\begin{align} \mathcal L_{GS} = \frac{1}{8}\text{Im}\left(f(s) \right) \ \epsilon^{\mu \nu \rho \sigma} F_{\mu \nu} F_{\rho \sigma}, \end{align}
and a gauge transformation (\ref{shift}) of the gauge kinetic function $f(s) = 1 + \beta_R s$ gives a contribution
\begin{align} \delta \mathcal L_{GS} =  -\theta \frac{\beta_R c}{8}   \ \epsilon^{\mu \nu \rho \sigma} F_{\mu \nu} F_{\rho \sigma}. \end{align}
The theory can be made gauge invariant by choosing
\begin{align} \beta_R = - \frac{ \mathcal C_R}{12 \pi^2 c} = -\frac{35 b^3 c^2}{96 \pi^2}. \end{align}

  \subsection{The mixed anomalies} \label{Appendix:mixed}
  We now calculate the cancellation conditions of the mixed anomalies by a Green-Schwarz mechanism.   In a theory with a gauged R-symmetry, the anomalous contributions to the triangle diagrams involving the R-current and two gauge fields or gravitons are given by
  \begin{align}   (F \tilde F)_A:& &  \mathcal C_A \delta^{ab} &= \Tr \left[ R_\psi (\tau^a \tau^b )_A \right] + T_{G_A} \delta^{ab} R_\lambda  \notag \\   \mathcal R \mathcal {\tilde R}: & &C_{\text{grav}} &= \Tr \left[ R_{\psi} \right] + n_\lambda R_\lambda - 21 R_{ \psi_{3/2} }.   \end{align}
  Here, $T_{G_A}\delta^{ab} = f^{acd} f^{bcd}$ with $T_{G_A} = N$ for $SU(N)$ and 0 for $U(1)$, $A$ labels the groups $U(1)_Y, SU(2)_L, SU(3)$.    The contribution of the gravitino is $-21$ times the contribution of a gaugino.    We can now calculate the $U(1)_R \times U(1)_Y^2$ anomaly   
  \begin{align}   \mathcal C_1 &= 3 \left[  \frac{\xi}{2}  \left(\frac{1}{6}  + \frac{4}{3} + \frac{1}{3} + \frac{1}{2} +1 \right)  \right] + \left(\frac{\xi}{2} \right) \left(\frac{1}{2} + \frac{1}{2} \right) \notag \\   &=  11 \left( \frac{\xi}{2} \right), \label{C1}   \end{align}
  the mixed $U(1)_R \times SU(2)$ anomaly
  \begin{align}   \mathcal C_2 &= \frac{3}{2} \left[  \left( \frac{\xi}{2} \right) \left(3 +1  \right) \right] + \frac{1}{2} \left( \frac{\xi}{2} \right) \left(1 + 1 \right) - 2 \left( \frac{\xi}{2} \right)  \notag \\   &=  5 \left( \frac{\xi}{2} \right), \label{C2}   \end{align}
  the mixed $U(1)_R \times SU(3)$ anomaly
  \begin{align}   \mathcal C_3 &= \frac{3}{2} \left[ \left( \frac{\xi}{2} \right)  \left(2+2 \right) \right] - 3 \left( \frac{\xi}{2} \right) \notag \\   &= 3 \left( \frac{\xi}{2} \right), \label{C3}   \end{align}
  and the gravitational anomaly
  \begin{align}   \mathcal C_{\text{grav}} &= 3 \left[ \left( \frac{\xi}{2} \right) \left(6 + 3 + 3 +2 + 1  \right) \right] + \left(\frac{\xi}{2} \right) \left(2 +2 +1 +1 +21 - 13  \right) \notag \\   &=  59 \left(\frac{\xi}{2} \right) .\label{Cgrav}   \end{align}
  In the equations above, the term in square brackets comes from the contributions of quarks and leptons $Q,u,d,L$ and $e$. The second term in the first line in eqs. (\ref{C1}) and (\ref{C2}) comes from the Higgsinos,    and the last terms in the first line of eqs. (\ref{C2}) and (\ref{C3}) is the contribution of the gauginos ($T_G$).   The contributions to the second term in the first line of eq. (\ref{Cgrav}) come from the Higgsinos, $\chi_s$,$\chi_z$, the gravitino and the gauginos respectively,    where $\chi_s$ and $\chi_z$ are the superpartners of $s$ and $z$ and we have $13 = 8 + 3 + 1 +1$ gauginos. In the above expressions, we used that $T_R = 11$ for $U(1)_Y$, $T_R=7$ for $SU(2)$ and $T_R=6$ for $SU(3)$.
  
  These anomalies are cancelled by a Green-Schwarz mechanism\footnote{The inclusion of appropriate counter terms that cancel the $G^2 \times U(1)_R$ mixed non-abelian anomaly and bring the abelian mixed anomaly to a covariant form are included in these results; for more information see \cite{FreedmanAnomalies}.} 
  \begin{align}   \mathcal L_{GS} = \frac{1}{8} \text{Im}( s) \epsilon^{\mu \nu \rho \sigma} \left( \beta_A F_{\mu \nu}^A F_{\rho \sigma}^A + \beta_{\text{grav}} \mathcal R_{\mu \nu} \mathcal {\tilde R}_{\rho \sigma}   \right),   \end{align}
  provided
  \begin{align}   \mathcal C_A &= - 4 \pi^2 c \ \beta_A , \ \ \ \ \ \ \ A=1,2,3 \notag \\   \mathcal C_{\text{grav}}  &=  32 \pi^2 c \ \beta_{\text{grav}}.   \end{align}
  This gives the anomaly cancellation conditions
  \begin{align}   \beta_1 &= - \frac{  11 \left(\xi/2 \right)}{4 \pi^2 c}, \notag \\   \beta_2 &= - \frac{ 5 \left(\xi/2 \right)}{4 \pi^2 c}, \notag \\   \beta_3 &= - \frac{3 \left(\xi/2 \right)}{4 \pi^2 c}. \label{anomalieformules}   \end{align}
   
   \vfill \end{document}